# A Hierarchical Bayesian Framework for Model-based Prognostics

Xinyu Jia[1*], Iason Papaioannou[1], Daniel Straub[1]

[1] *Engineering Risk Analysis Group, Technische Universität München, Arcisstr. 21, 80290 München, Germany*

## ABSTRACT

In prognostics and health management (PHM) of engineered systems, maintenance decisions ideally are informed by predictions of a system's remaining useful life (RUL) based on operational data. Model-based prognostics algorithms rely on a parametric model of the system degradation process. The model parameters are learned from real-time operational data collected on the system. However, there can be valuable information in data from similar systems or components, which is not typically utilized in PHM. In this contribution, we propose a hierarchical Bayesian modelling (HBM) framework for PHM that integrates both operational data and run-to-failure data from similar systems or components. The HBM framework utilizes hyperparameter distributions learned from data of similar systems or components as priors. It enables an efficient update of predictions as more information becomes available, allowing for an increasingly accurate assessment of the degradation process and its associated variability. The effectiveness of the proposed framework is demonstrated through two experimental applications, which involve real-world data from crack growth and lithium batteries degradation. Results show significant improvements in RUL prediction accuracy and demonstrate how the framework facilitates the management of uncertainty through predictive distributions.

## 1. Introduction

Prognostics and health management (PHM) utilizes monitoring data to assess and predict the health status of engineering systems and to optimize maintenance planning [1,2]. A key quantity of interest in PHM is the remaining useful life (RUL) of a component or system. The RUL is the system lifetime left from the current time or load-cycle to the end-of-life (EOL), with the EOL representing the time at which the system health crosses a failure threshold [3]. Prognostics refers to the online prediction of the RUL using real-time and historical monitoring data. By accurately estimating RUL, PHM enables effective management of uncertainty, ensures timely interventions, and improves overall system reliability and lifecycle management.

Prognostics methods can be classified into data-driven and physics-based methods [4,5]. Data-driven methods rely on processing and analyzing historical operational data through regression and classification. Linear basis function models [6], support vector machines [7], Gaussian process regression [8,9], and neural networks [10–12] have been successfully applied to prognostics. The primary advantage of data-driven methods is their ability to handle complex and nonlinear relationships in degradation processes without requiring explicit knowledge of the underlying physical processes. However, data-driven prognostics algorithms are highly dependent on the availability and quality of training data and often struggle to generalize beyond the conditions they were trained on. Additionally, these methods can lack model interpretability in practice. One approach to address these issues is physics-informed machine learning [13–15], which combines data-driven modeling with physical principles, allowing for better interpretability and broader applicability. However, these methods require a high-fidelity physical model to be embedded into the machine learning framework [16,17]. A review of physics-informed machine learning for PHM of batteries is provided in [18].

[*] Corresponding author.
E-mail address: xinyujia@hebut.edu.cn

Furthermore, uncertainty quantification has also been a focus in data-driven approaches, enabling more robust predictions, as reviewed in [19].

By contrast, model-based approaches predict the RUL by fitting empirical or physical parametric models that describe the degradation process. A major advantage of model-based methods is their reliance on physics-based principles, and as a result, fewer training data is required for accurate predictions [20]. Model-based prognostics often apply Bayesian algorithms, which offer a probabilistic framework to update the model parameters and RUL predictions dynamically as new data becomes available [21]. For instance, Zio and Peloni [22] propose a methodology for the estimation of the remaining useful life of components based on particle filtering. An et al. [23] propose a Bayesian statistical methodology for estimating the wear coefficient and predicting wear volume in a revolute joint using in-situ measurement data.

When data is noisy or shows irregular patterns, especially in the early stages of degradation, a model-based approach may result in inaccurate RUL predictions. To address this issue, Jun et al. [24] incorporated physical constraints within the Bayesian framework to improve the accuracy of RUL predictions. Although model-based methods offer interpretability and robustness, their performance heavily depends on the availability of accurate prior knowledge about the system before incorporating system-specific data. Accurate prior knowledge can significantly enhance the reliability of predictions by capturing the inherent variability and uncertainty in the degradation process. Historical data, whether from similar components or different components of the same system, can provide valuable prior information that can improve the accuracy of RUL predictions by reflecting trends and variability observed in past degradation processes. Uncertainty in RUL predictions arises from various factors, such as environmental conditions, manufacturing processes, and material properties, and much of this uncertainty is inherently captured in historical data. Inadequately chosen priors or neglecting historical data can result in inaccurate predictions, particularly in systems with complex or poorly understood failure mechanisms. In such cases, even the most advanced model-based approaches may struggle to capture the true degradation process.

Motivated by these challenges, this paper presents a hierarchical Bayesian modeling (HBM) framework for prognostics and RUL prediction. The proposed framework addresses the gap in the field of model-based prognostics by integrating historical data from similar components or other components within the same system to enhance degradation predictions. It allows the model to leverage valuable historical data to enhance predictions for the current component, while accounting for uncertainties in material properties, environmental conditions, and operational variability. Although HBM has gained attention in fields such as structural dynamics [25–27], geotechnical engineering [28–30], and molecular dynamics [31–33] for addressing multi-source uncertainties, its application to PHM remains scarce, with limited studies exploring its potential in battery prognostics [34,35]. Xu et al. [34] combine discharging and degradation models to predict the end-of-discharge cycles and remaining useful cycles. Mishra et al. [35] apply the framework to analyze and predict battery discharge behavior under variable load profiles. These studies are tailored to specific applications and do not offer a general framework for integrating historical data into the RUL prediction of a specific component. Our work addresses this gap by establishing a generalizable HBM framework that incorporates historical data as prior information to enhance RUL predictions for the current component. It accounts for uncertainties from historical datasets, enabling a more accurate and robust assessment of degradation evolution. By combining historical and current data, the proposed framework provides a robust and versatile solution for PHM, advancing the accuracy of RUL predictions in complex systems.

The remainder of this paper is organized as follows: Section 2 introduces the proposed HBM framework and its mathematical formulations for PHM. Section 3 investigates the application of the framework to a fatigue crack growth problem, while Section 4 applies it to a battery degradation process. Finally, Section 5 presents conclusions and discusses potential directions for future work.

## 2. An HBM framework for PHM

### *2.1 Problem statement*

Model-based prognostics employs a degradation model $g(\boldsymbol{\theta}, t)$ with model parameters $\boldsymbol{\theta}$ to predict the system state at time $t$. Such predictions are subject to uncertainties due to variability in experimental tests, environmental conditions, manufacturing variabilities, material property randomness and model errors. The Bayesian approach systematically accounts for these uncertainties by providing the posterior predictive probability distribution of future degradation states. To achieve this, the model parameters of a system of interest are inferred based on data collected until the current time or load cycle via Bayes' rule, which defines the posterior distribution of the parameters as a product of the data likelihood and prior distribution. The EOL of the system is defined as the time when the degradation state reaches a critical threshold, and the system has to be maintained. The RUL is the difference between the EOL and the current time. In the Bayesian approach, the predicted RUL is a random variable, as it depends on the model parameters described by the Bayesian posterior distribution. Maintenance planning is based on either the full distribution or a conservative estimate – typically a lower quantile of the RUL.

The classical Bayesian approach to RUL prediction does not utilize data from similar components/systems. However, such historical (or fleet) data can be used to inform the prior distribution of the current component's parameters. The proposed HBM framework learns the posterior distribution of the component-specific model parameters by combining historical data with component-specific data. The detailed mathematical formulation of the proposed HBM framework is introduced in the following sections.

The proposed approach can be applied to components or systems, but to simplify the presentation, we will only use the term component in the following.

### *2.2 The HBM model*

Figure 1 summarizes the proposed HBM framework for model-based prognostics. The framework includes a training and a learning process, where the training process involves the historical data, and the learning process contains the component-specific data. $\mathbf{D}_h = \{\mathbf{D}_i, i = 1, 2, \cdots, N\}$ denotes the historical data, where $N$ is the number of datasets in the training process, and $\mathbf{D}_c$ denotes the degradation data up to the current time or load cycle of the component of interest for which the RUL is to be predicted. Historical data generally represents degradation data until a failure threshold is reached, taken from components similar to the component of interest. For example, the historical data could be from the same type of aircraft, or from different panels on the same aircraft, which implies that the usage conditions underlying the historical data may vary from those in the component of interest.

Each historical dataset $\mathbf{D}_i$ is assumed to correspond to a specific set of model parameters $\boldsymbol{\theta}_i$, which can be inferred using the corresponding dataset $\mathbf{D}_i$ through Bayesian analysis. The goal is to predict the future behavior of degradation and the RUL of the component of interest through learning the model parameters related to this component $\boldsymbol{\theta}_c$ based on the historical data $\mathbf{D}_h$ and component-specific data $\mathbf{D}_c$.

The model parameters $\boldsymbol{\theta}_i$, $i = 1, 2, \cdots, N$, learned from historical data, along with the parameters $\boldsymbol{\theta}_c$ of the current component, are assumed to belong to the same statistical population. The variability within this population is captured by introducing hyperparameters $\boldsymbol{\psi}$ that define the probability distribution of the model parameters across the components in the population. Historical data $\mathbf{D}_h$ can be used to learn an informative prior of $\boldsymbol{\psi}$ for the inference of the parameters of the current component.

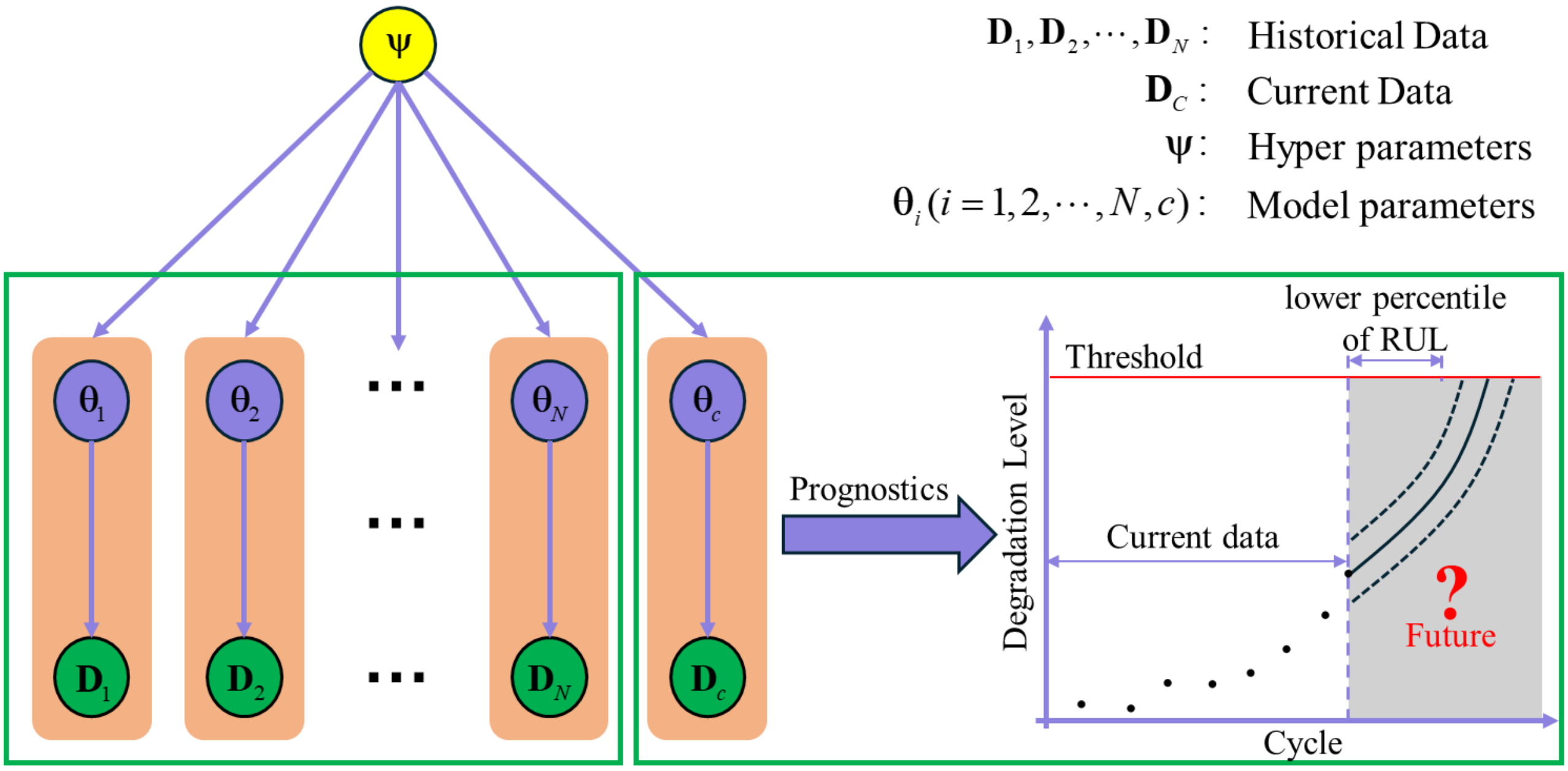

Figure 1. The proposed HBM framework for PHM

The conditional probability density function (pdf) $p(\boldsymbol{\theta}_i \mid \boldsymbol{\psi})$ is assumed to follow a Gaussian distribution [36,37], i.e.,

$$p(\boldsymbol{\theta}_i \mid \boldsymbol{\psi}) = N(\boldsymbol{\theta}_i \mid \boldsymbol{\mu}_{\boldsymbol{\theta}}, \boldsymbol{\Sigma}_{\boldsymbol{\theta}}), \tag{1}$$

with hyperparameters $\boldsymbol{\psi}$ = [$\boldsymbol{\mu}_{\boldsymbol{\theta}}$, $\boldsymbol{\Sigma}_{\boldsymbol{\theta}}$] representing the mean $\boldsymbol{\mu}_{\boldsymbol{\theta}}$ and covariance matrix $\boldsymbol{\Sigma}_{\boldsymbol{\theta}}$, and $i = 1, 2, \ldots, N, c$. The Gaussian distribution model is motivated by the principle of maximum entropy [38]. By selecting a Gaussian hierarchical prior, the model allows for the incorporation of known information, such as mean and covariance, while maintaining flexibility in capturing the underlying variability in the historical datasets. Given the similarity between the current component and historical components, their corresponding model parameters are assumed to follow the same Gaussian hierarchical structure, facilitating knowledge transfer across components. This structure not only ensures coherence between components but also allows the model to dynamically update its predictions as new data from the current component are incorporated. Once the posterior distribution of the model parameters in the current component $\boldsymbol{\theta}_c$ given the historical and current data is obtained, the future behavior of degradation and the RUL of the component can be predicted. Detailed formulations for the inference of the hyperparameter, the model parameters as well as the degradation prognostics and RUL prediction are given in Section 2.3.

### *2.3 Inference and prognostics with the HBM framework*

#### *2.3.1 Learning the hyperparameters using historical data*

In the initial training process, the historical data $\mathbf{D}_h$ are used to infer the hyperparameter distribution that then serves as the prior for the model parameters of the component of interest. The posterior pdf of the hyperparameters $p(\boldsymbol{\psi} \mid \mathbf{D}_h)$ is [39]

$$p(\boldsymbol{\psi} \mid \mathbf{D}_h) = \frac{p(\mathbf{D}_h \mid \boldsymbol{\psi})\, p(\boldsymbol{\psi})}{p(\mathbf{D}_h)}, \tag{2}$$

where $p(\boldsymbol{\psi})$ is the prior pdf of the hyperparameters, $p(\mathbf{D}_h)$ is the model evidence and $p(\mathbf{D}_h \mid \boldsymbol{\psi})$ is the likelihood function that represents the probability of observing the given datasets $\mathbf{D}_h$ conditional on hyperparameter values $\boldsymbol{\psi}$. Assuming independence across the historical datasets, the likelihood function $p\left(\mathbf{D}_h|\boldsymbol{\psi}\right)$ is expressed as the product of each likelihood function given a specific historical dataset $\mathbf{D}_i$:

$$p\left(\mathbf{D}_h|\boldsymbol{\psi}\right)=\prod_{i=1}^{N} p\left(\mathbf{D}_i|\boldsymbol{\psi}\right). \tag{3}$$

Following the model of Fig. 1, the likelihood function $p\left(\mathbf{D}_i|\boldsymbol{\psi}\right)$ is written as

$$p\left(\mathbf{D}_i|\boldsymbol{\psi}\right)=\int p\left(\mathbf{D}_i|\boldsymbol{\theta}_i\right)p\left(\boldsymbol{\theta}_i|\boldsymbol{\psi}\right)\mathrm{d}\boldsymbol{\theta}_i, \tag{4}$$

where $p\left(\boldsymbol{\theta}_i|\boldsymbol{\psi}\right)$ is the hierarchical prior defined in Eq. (1), and $p\left(\mathbf{D}_i|\boldsymbol{\theta}_i\right)$ is the likelihood function for each dataset $\mathbf{D}_i$ given the model parameters $\boldsymbol{\theta}_i$. The likelihood function $p\left(\mathbf{D}_i|\boldsymbol{\theta}_i\right)$ models the discrepancy between the model predictions and the measurements. Specific likelihood functions for the crack growth and battery degradation applications are given in Sections 3 and 4.

The integral in Eq. (4) can be solved via the Laplace approximation, which approximates the likelihood function $p\left(\mathbf{D}_i|\boldsymbol{\theta}_i\right)$ around its mode by a Gaussian density [40]. This approach is particularly effective when the $i$th data set contains many data points, in which case the Gaussian approximation is more accurate. With the Laplace approximation and the Gaussian parameter prior of Eq. (1), the integral of Eq. (4) can be evaluated in closed form. This method has been shown to be effective in structural dynamics, where it is applied to analyze modal properties [27] and time history data [36]. However, data is often limited in PHM within a dataset, making the Laplace approximation less suitable.

Alternatively, the integral of Eq. (4) can be approximated by Monte Carlo (MC) methods, through generating samples from the posterior distribution resulting from an auxiliary Bayesian updating with likelihood $p\left(\mathbf{D}_i|\boldsymbol{\theta}_i\right)$ and uniform prior [41]. This approach is flexible as it does not require assumptions on the likelihood shape and can handle complex integrands. Using this approximation, Eq. (4) takes the following form:

$$p\left(\mathbf{D}_i|\boldsymbol{\psi}\right)\approx\frac{1}{N_k}\sum_{k=1}^{N_k} p\left(\boldsymbol{\theta}_i^{(k)}|\boldsymbol{\psi}\right), \tag{5}$$

where $\boldsymbol{\theta}_i^{(k)}$ is the $k$-th sample drawn from the distribution proportional to $p\left(\mathbf{D}_i|\boldsymbol{\theta}_i\right)$ and $N_k$ is the total number of Monte Carlo samples used for integration. The samples $\boldsymbol{\theta}_i^{(k)}$, $k = 1,\ldots,N_k$ can be generated by any sampling algorithm capable of sampling distributions known up to a normalizing constant, such as Markov chain Monte Carlo methods. Substituting Eq. (5) into Eq. (2) leads to:

$$p\left(\boldsymbol{\psi}|\mathbf{D}_h\right)\propto p\left(\boldsymbol{\psi}\right)\prod_{i=1}^{N}\frac{1}{N_k}\sum_{k=1}^{N_k} p\left(\boldsymbol{\theta}_i^{(k)}|\boldsymbol{\psi}\right)\propto p\left(\boldsymbol{\mu}_{\boldsymbol{\theta}},\boldsymbol{\Sigma}_{\boldsymbol{\theta}}\right)\prod_{i=1}^{N}\sum_{k=1}^{N_k} N\left(\theta_i^{(k)}\,|\,\boldsymbol{\mu}_{\boldsymbol{\theta}},\boldsymbol{\Sigma}_{\boldsymbol{\theta}}\right). \tag{6}$$

The posterior of Eq. (6) can be approximated by any available Bayesian updating algorithm. As a result, the computation for the posterior distribution of the hyperparameters given the historical dataset, $p\left(\boldsymbol{\psi}|\mathbf{D}_h\right)$, is performed into two steps. In the first step, one needs to perform a classical Bayesian inference for each dataset $\mathbf{D}_i$ to obtain the posterior samples of model parameters $\boldsymbol{\theta}_i$ assuming a uniform prior on $\boldsymbol{\theta}_i$. The obtained samples are directly used in the second step to compute the posterior distribution of the hyperparameters using Eq. (6). To facilitate implementation, a pseudocode for computing the posterior distribution of the hyperparameters is given in Appendix A.

### 2.3.2 Learning the model parameters using current data

The data from the current component $\mathbf{D}_c$ is employed to infer the model parameters $\boldsymbol{\theta}_c$, while accounting for information from historical data. This is achieved through the shared hierarchical structure of the model parameters of similar components described by the prior of Eq. (1). From Bayes' rule, the posterior distribution of the model parameters given all data follows as:

$$p\left(\boldsymbol{\theta}_c \middle| \mathbf{D}_c, \mathbf{D}_h\right) \propto p\left(\mathbf{D}_c \middle| \boldsymbol{\theta}_c\right) p\left(\boldsymbol{\theta}_c \middle| \mathbf{D}_h\right), \tag{7}$$

where $p\left(\mathbf{D}_c \middle| \boldsymbol{\theta}_c\right)$ is the likelihood function corresponding to the current data, and $p\left(\boldsymbol{\theta}_c \middle| \mathbf{D}_h\right)$ is the prior pdf of the model parameters $\boldsymbol{\theta}_c$ informed by the historical data $\mathbf{D}_h$. With the model of Fig. 1, it is determined as:

$$p\left(\boldsymbol{\theta}_c \middle| \mathbf{D}_h\right) = \int p\left(\boldsymbol{\theta}_c \middle| \boldsymbol{\psi}\right) p\left(\boldsymbol{\psi} \middle| \mathbf{D}_h\right) \mathrm{d}\boldsymbol{\psi}. \tag{8}$$

The integral in Eq. (8) can be approximated using $N_s$ MC samples of the hyperparameters $\boldsymbol{\psi}^{(s)}$ drawn from the conditional distribution $p\left(\boldsymbol{\psi} \middle| \mathbf{D}_h\right)$ according to Eq. (6):

$$p\left(\boldsymbol{\theta}_c \middle| \mathbf{D}_h\right) \approx \frac{1}{N_s} \sum_{s=1}^{N_s} p\left(\boldsymbol{\theta}_c \middle| \boldsymbol{\psi}^{(s)}\right). \tag{9}$$

Substituting Eq. (9) into Eq. (7) leads to:

$$p\left(\boldsymbol{\theta}_c \middle| \mathbf{D}_c, \mathbf{D}_h\right) \propto p\left(\mathbf{D}_c \middle| \boldsymbol{\theta}_c\right) \frac{1}{N_s} \sum_{s=1}^{N_s} p\left(\boldsymbol{\theta}_c \middle| \boldsymbol{\psi}^{(s)}\right). \tag{10}$$

$p\left(\boldsymbol{\theta}_c \middle| \mathbf{D}_c, \mathbf{D}_h\right)$ is the posterior PDF when using all the datasets. This expression highlights the contrast between using only current component-specific data and incorporating both historical datasets and current data. In particular, the term $\frac{1}{N_s} \sum_{s=1}^{N_s} p\left(\boldsymbol{\theta}_c \middle| \boldsymbol{\psi}^{(s)}\right)$ in Eq. (10) serves as a prior distribution for the model parameter $\boldsymbol{\theta}_c$, derived from the information in the historical datasets. In the absence of historical data, priors are typically chosen to be uninformative or weakly informative. Using informative priors based on historical data is expected to result in decreased posterior uncertainty. Moreover, it avoids arbitrary prior choices that can potentially lead to biased posterior estimates.

### 2.3.3 Degradation prognostics and RUL prediction

The posterior distribution of the model parameter $\boldsymbol{\theta}_c$ can be used to perform degradation prognostics. Let $\boldsymbol{\theta}_c^{(l)}$ denote the $l$-th sample from the posterior distribution $p\left(\boldsymbol{\theta}_c \middle| \mathbf{D}_c, \mathbf{D}_h\right)$. The corresponding $l$-th prediction of the degradation at cycle $t$ is:

$$\mathbf{y}_t^{(l)} = g(\boldsymbol{\theta}_c^{(l)}, t). \tag{11}$$

The simulation results in $N_l$ samples of the degradation $\mathbf{y}_t$. The statistics, including quantile estimates as well as the uncertainty bounds of the model predictions can be easily obtained based on these samples.

When the predicted degradation reaches a prescribed failure threshold $g_t$, the EOL is reached, denoted by $t_{EOL}^{(l)}$. The RUL is then predicted by subtracting the current cycle $t_c$ from the EOL:

$$t_{RUL}^{(l)} = t_{EOL}^{(l)} - t_c. \tag{12}$$

This results in $N_l$ samples of the RUL. The corresponding empirical RUL prediction distribution can then be used as a basis for maintenance optimization [42–44].

## 3. Application to Crack Growth

### *3.1 Experimental datasets and crack growth model*

The proposed HBM framework for prognostics and RUL prediction is validated using public datasets provided by NASA's Open Data Portal [45–47]. Specifically, we consider fatigue experiments that were conducted on aluminum lap-joint specimens, in which lamb signals were recorded for each specimen at regular number of cycles during fatigue testing. The signals from piezo actuator-receiver sensor pairs were reported and these signals were directly related to the fatigue crack lengths. The geometry of the lap-joint component and the sensor network configuration are shown in Figure 2. More details on the data generating process, experimental setting and testing protocols can be found in [46]. With the above setup, eight experimental datasets (named T1-T8) were obtained; they are plotted in Figure 3. The first seven specimens (T1-T7) were tested under constant loading while the last specimen (T8) was tested under varied amplitude loading. For the purpose of our investigation, the eight specimens are partitioned into historical data and current components for prediction. T1–T6 provide historical datasets for inferring the HBM, while T7–T8 are treated as the current components whose RUL is predicted based on the inferred model and then compared with their subsequent measurements.

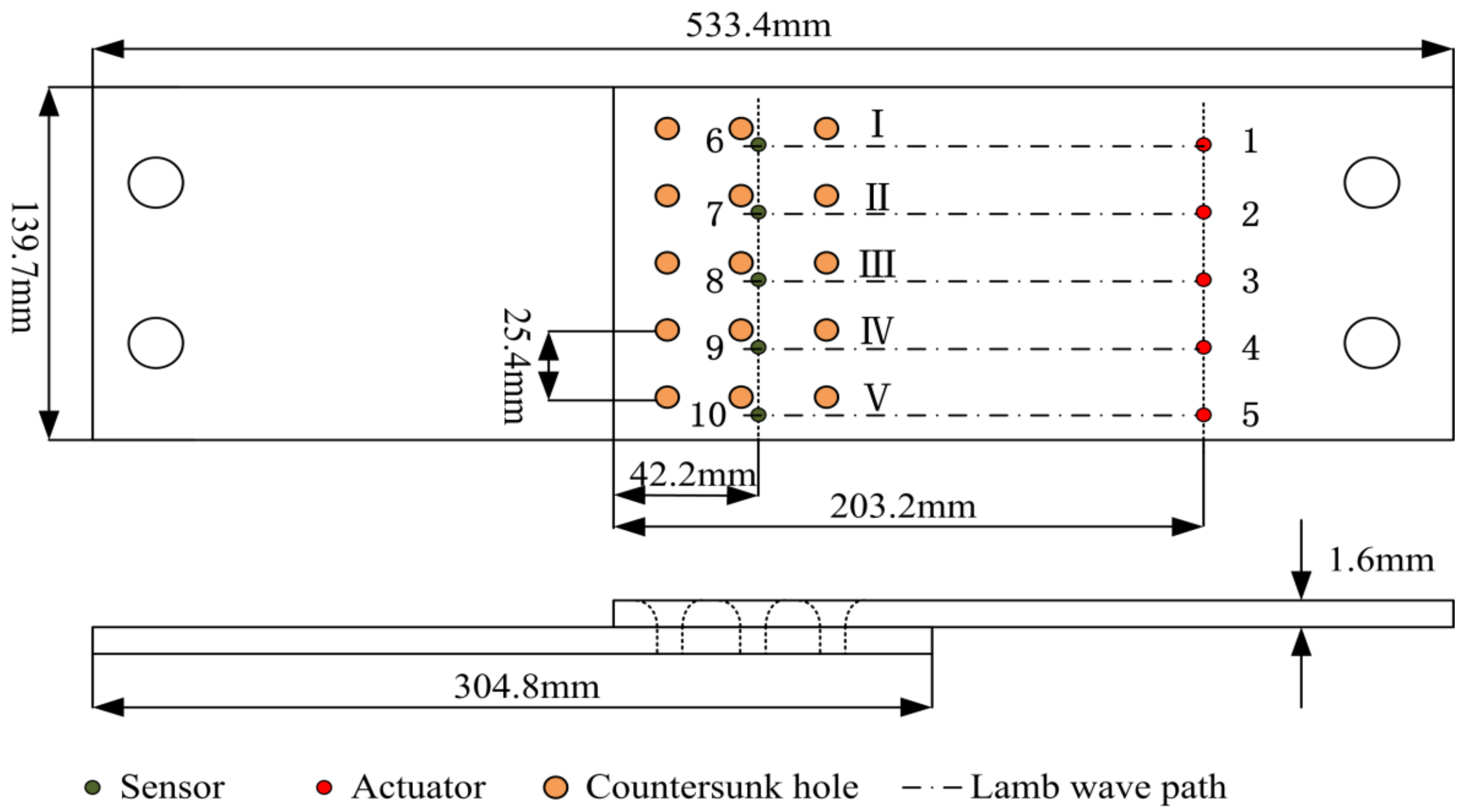

Figure 2. The geometry of the lap-joint component and sensor network configuration for fatigue testing [46].

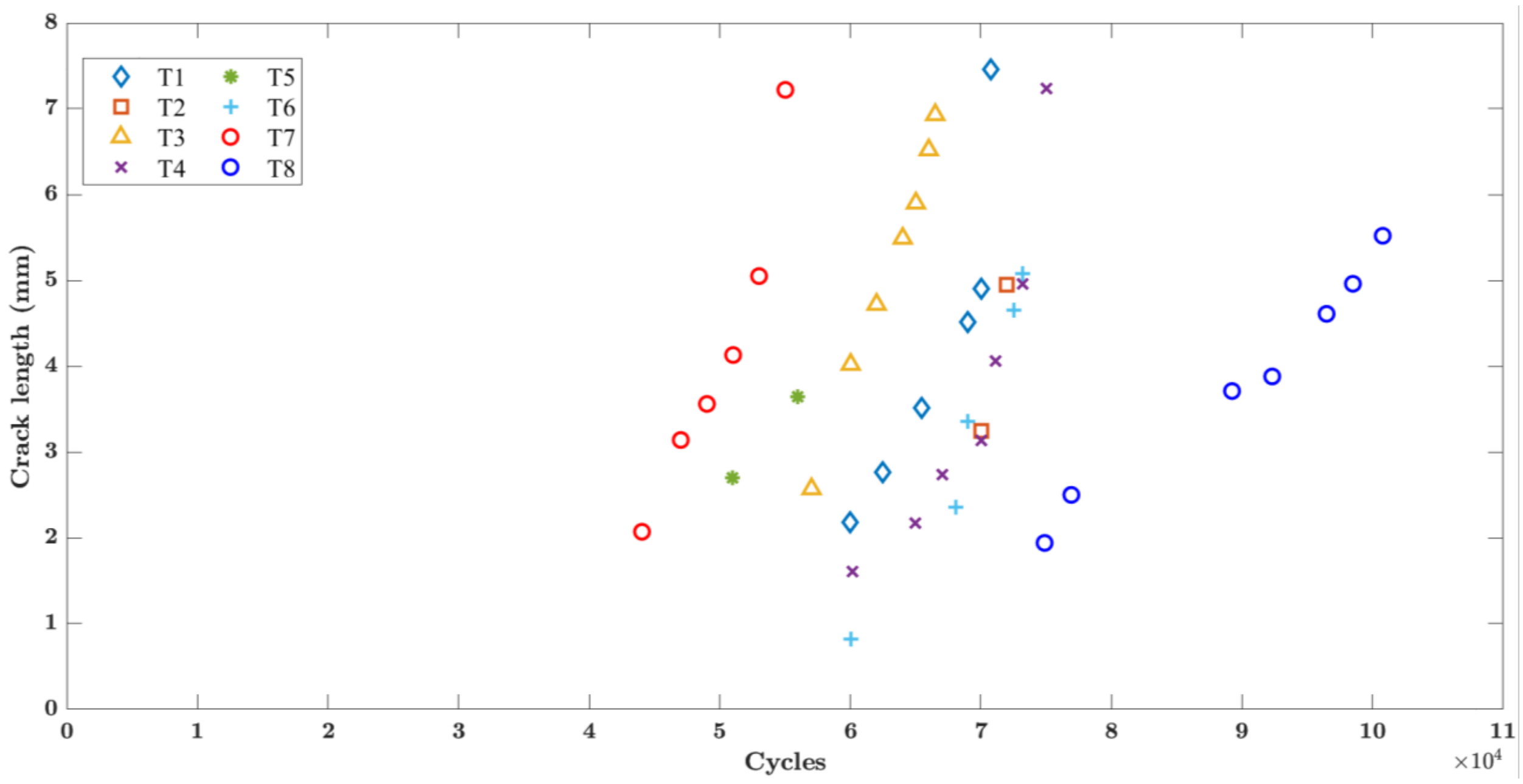

Figure 3. Experimental tests (T1-T8)

This experimental dataset has been explored previously for degradation prognostics using data-driven and model-based approaches. We utilize the crack growth model of [48,49], where the Paris' law is adopted for the evaluation of the crack length for given loading cycles [50]:

$$\frac{da}{dN} = C\left(\Delta K\right)^m, \tag{13}$$

where $a$ is the crack length, $N$ is the number of cycles, $\Delta K = \Delta\sigma\sqrt{\pi a}$ is the range of the stress intensity factor with the applied stress amplitude $\Delta\sigma$, $C$ and $m$ are the empirically determined parameters in Paris' law. In this work, the parameters $C$ and $m$ are treated as unknown quantities and are inferred from data within the proposed HBM framework. For a given specimen under fixed loading conditions, $C$ and $m$ are assumed constant with respect to the cycle count $N$. For variable-amplitude two-block loading (used for specimen T8), we compute an equivalent stress amplitude $\Delta\sigma_{eq}$ by matching the cycle-averaged Paris driving term $\left(\Delta\sigma\right)^m$ [50]:

$$\Delta\sigma_{eq} = \left(\frac{N_1\left(\Delta\sigma_1\right)^m + N_2\left(\Delta\sigma_2\right)^m}{N_1 + N_2}\right)^{\frac{1}{m}}, \tag{14}$$

where $\Delta\sigma_1$ and $\Delta\sigma_2$ are the stress amplitudes in the two constant-amplitude blocks, and $N_1$ and $N_2$ are the two corresponding constant block loadings. Starting from a crack with length $a_0$ at $N_0$ cycles, the crack length $a$ after $N$ cycles is obtained as [49]:

$$a = \left(a_0^{1-\frac{m}{2}} + \left(1-\frac{m}{2}\right)\left(N-N_0\right)C\left(\Delta\sigma\sqrt{\pi}\right)^m\right)^{\frac{2}{2-m}}, \tag{15}$$

By setting the crack length to the critical length $a = a_f$ and solving Eq. (15) for $N$ gives the number of cycles to failure $N_f$:

$$N_f = N_0 + \frac{a_f^{1-\frac{m}{2}} - a_0^{1-\frac{m}{2}}}{\left(1-\frac{m}{2}\right)C\left(\Delta\sigma\sqrt{\pi}\right)^m}. \tag{16}$$

The predicted remaining useful life, in cycles, is then computed as $RUL = N_f - N_0$.

### *3.2 Setup of the numerical investigation*

With the experimental data and the model above, we estimate the Paris parameters $m$ and $C$ (in log-scale for $C$) together with the additive error standard deviation $\sigma$ in Bayesian inference. To improve numerical stability and to reduce scale differences between the parameters, we work with normalized parameters defined relative to their nominal physical values. Specifically, we introduce nominal values $m_0 = 2$ and $\log(C_0) = -18.6$ (taken from [49]) and re-parameterize the unknowns in a dimensionless form: $\theta_1 = m / m_0$, and $\theta_2 = \log(C)/\log(C_0)$. The likelihood function for this case is then assigned to be a lognormal distribution given the $i$-th dataset at $k$-th cycle $y_{i,k}$ [51]:

$$p\left(y_{i,k} \middle| \boldsymbol{\theta}_i\right) = \frac{1}{y_{i,k}\sqrt{2\pi}\zeta_{i,k}} \exp[-\frac{1}{2}(\frac{\ln y_{i,k} - \eta_{i,k}}{\zeta_{i,k}})^2], \tag{17}$$

where $\zeta_{i,k} = \sqrt{\ln\left[1 + \left(\frac{\sigma}{a(\boldsymbol{\theta}_i, k)}\right)^2\right]}$ and $\eta_{i,k} = \ln\left[a(\boldsymbol{\theta}_i, k)\right] - \frac{1}{2}(\zeta_{i,k})^2$ are the parameters of the lognormal distribution and $a(\boldsymbol{\theta}, k)$ is determined according to Eq. (15).

The hyperparameters associated with the model parameter $\boldsymbol{\theta}$ are $\boldsymbol{\mu}_{\boldsymbol{\theta}} = [\mu_{\theta_1}, \mu_{\theta_2}]$ and $\boldsymbol{\Sigma}_{\boldsymbol{\theta}}$, where $\boldsymbol{\Sigma}_{\boldsymbol{\theta}}$ is the covariance matrix that accounts for parameter uncertainty and the correlation between model parameters. We investigate two different models (Case 1 and Case 2). Case 1 assumes a diagonal matrix with the diagonal elements $\sigma_{\theta_1}^2$ and $\sigma_{\theta_2}^2$, while Case 2 considers a full matrix with both the diagonal elements and a correlation coefficient $\rho$ between model parameters $\theta_1$ and $\theta_2$. The hyperparameters for the prediction standard deviation $\sigma$ are $\mu_\sigma$ and $\sigma_\sigma^2$, where a truncated Gaussian prior distribution with range (0, 0.2) is chosen for $\sigma$ to ensure a non-negative $\sigma$.

For the numerical investigations, the datasets T1-T6 are considered as the historical datasets. The posterior distributions of the hyperparameters are computed using the Algorithm given in Appendix A. The priors of the hyperparameters are summarized in Table 1. The slice sampler [52] is applied to generate samples of the posterior distributions of each individual Bayesian updating problem in the proposed algorithm, whereby a total of 5000 samples is used.

Table 1. Prior distributions of the hyperparameters

| | $\mu_{\theta_1}$ | $\mu_{\theta_2}$ | $\mu_\sigma$ | $\sigma_{\theta_1}$ | $\sigma_{\theta_2}$ | $\sigma_\sigma$ | $\rho$ |
|---|---|---|---|---|---|---|---|
| Case 1 | U(0.8,1.4) | U(0.9,1.4) | U(0,0.4) | U(0,0.3) | U(0,0.1) | U(0,0.2) | - |
| Case 2 | U(0.8,1.4) | U(0.9,1.4) | U(0,0.4) | U(0,0.3) | U(0,0.1) | U(0,0.2) | U(-1,1) |

### *3.3 Results*

#### *3.3.1 Historical posterior distribution of the hyperparameters (historical data only)*

To avoid ambiguity, we distinguish two posterior stages in this work. The first one, referred as historical posterior, is inferred from T1–T6 and is then used as an informative prior for T7/T8. Conditioning on the available measurements for either T7 or T8 then yields the updated posterior (the second stage) for the current component (Section 3.3.3).

Figures 4 and 5 show the hyperparameter distribution conditional on the historical data for the two modeling assumptions: Case 1 (no correlation) and Case 2 (with correlation). Table 2 also summarizes the first two moments (mean and standard deviation) of the inferred hyperparameters for both cases. Overall, the inferred hyperparameter means are similar between the two cases, indicating that both formulations lead to comparable central estimates when calibrated on T1–T6. We note that the correlation coefficient $\rho$ in Case 2 has a positive mean value, which at first glance might appear to contradict the fact that a negative correlation factor is commonly reported in the literature for parameters $m$ and $\ln C$. The reason lies in the fact that $\rho$ is the correlation coefficient of the normalized parameters $\theta_1 = m/m_0$ and $\theta_2 = \ln\ C/\ln\ C_0$, rather than that of the physical parameters themselves. Since the nominal value $\ln C_0$ is negative, this normalization leads to a sign change in the correlation, so that a negative correlation between $m$ and $\ln C$ results in a positive correlation between $\theta_1$ and $\theta_2$.

Table 2. Estimates of the statistics of the historical hyperparameters for the two cases

| | Statistics | $\mu_{\theta_1}$ | $\mu_{\theta_2}$ | $\mu_{\sigma}$ | $\sigma_{\theta_1}$ | $\sigma_{\theta_2}$ | $\sigma_{\sigma}$ | $\rho$ |
|---|---|---|---|---|---|---|---|---|
| Case 1 | Mean | 1.03 | 1.05 | 0.09 | 0.12 | 0.02 | 0.06 | - |
| | S.D. | 0.07 | 0.01 | 0.03 | 0.06 | 0.01 | 0.05 | - |
| Case 2 | Mean | 1.03 | 1.05 | 0.08 | 0.14 | 0.02 | 0.06 | 0.45 |
| | S.D. | 0.07 | 0.01 | 0.03 | 0.06 | 0.01 | 0.04 | 0.45 |

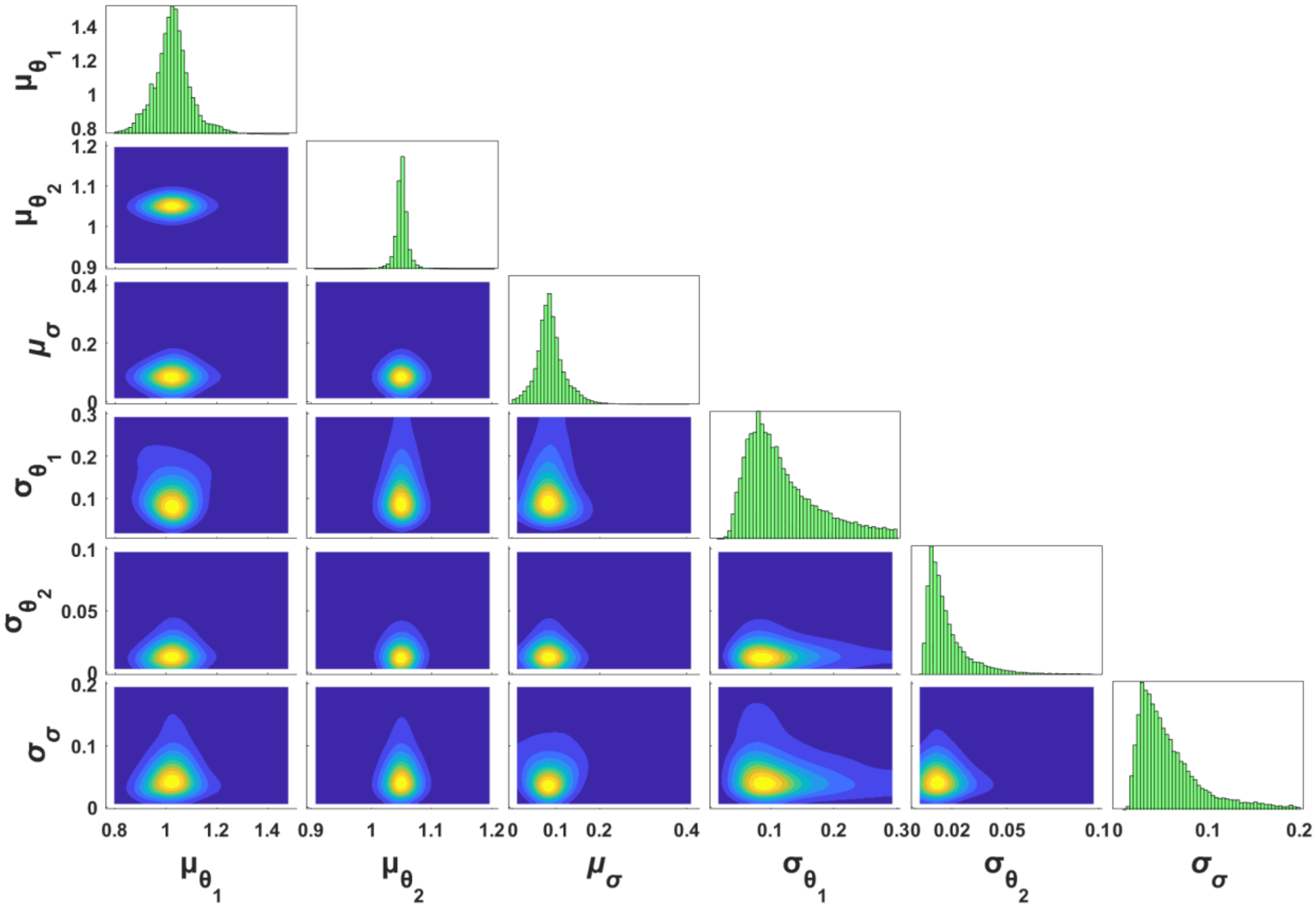

Figure 4. Historical posterior distributions of hyperparameters learned from the T1–T6 for Case 1 (independent parameters). The diagonals display histograms of the posterior samples, the off-diagonals show heatmaps of the bivariate posterior distributions.

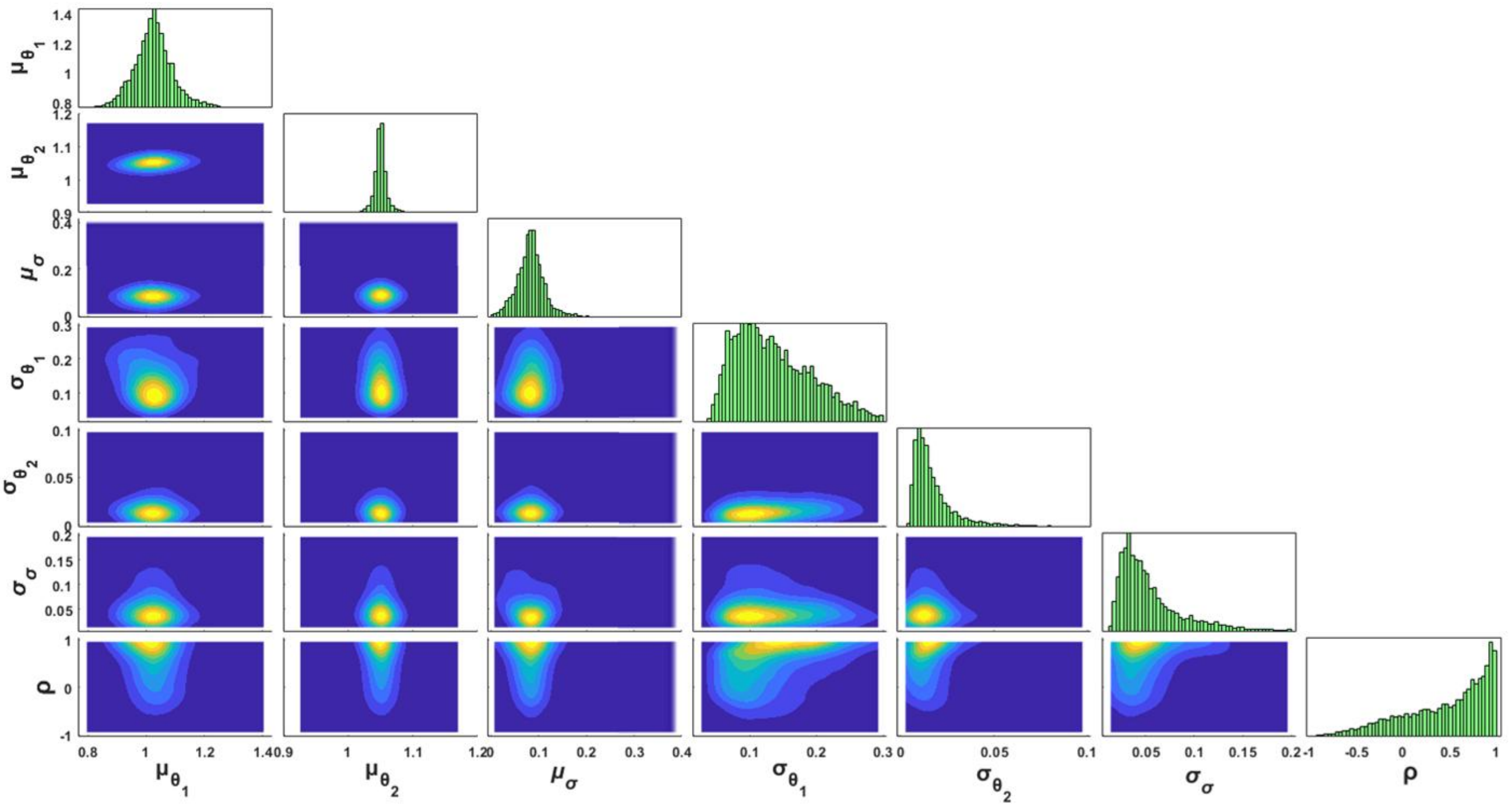

Figure 5. Historical posterior distributions of hyperparameters learned from the T1–T6 for Case 2 (correlated parameters). The diagonals display histograms of the posterior samples, the off-diagonals show heatmaps of the bivariate posterior distributions, and the last panel displays the posterior of the correlation between $\theta_1$ and $\theta_2$.

### *3.3.2 Posterior model parameter distribution and crack growth predictions based on historical data only*

Figure 6 shows the posterior statistics of the hyperparameters conditional on the historical datasets (T1–T6). The marginal posteriors of $\theta_1$, $\theta_2$, and $\sigma$ are broadly similar in both cases, while Case 2 additionally captures the dependence between $\theta_1$and $\theta_2$. We then propagate these parameter samples through the crack growth model to predict crack length (Figure 7). Both cases produce credible intervals that cover all experimental observations, however, including parameter correlation (Case 2) yields narrower credible intervals, indicating that accounting for correlation reduces predictive uncertainty compared with the independent parameter assumption (Case 1).

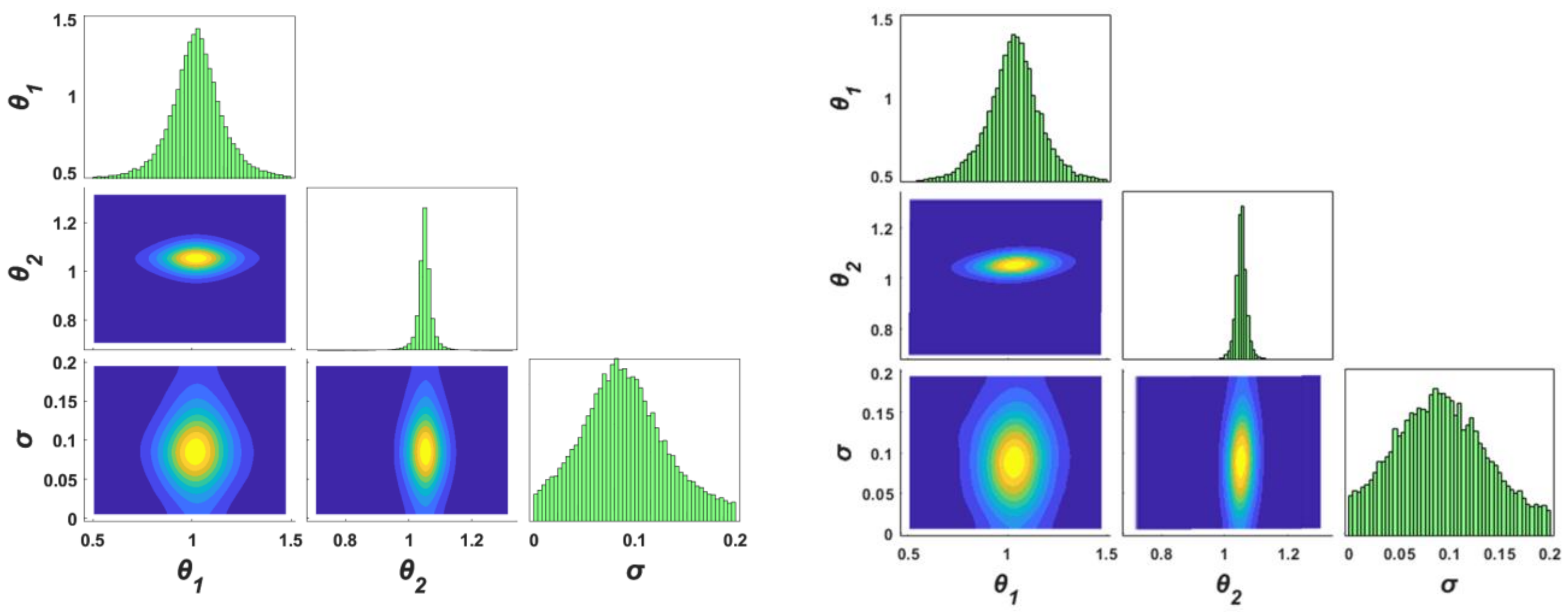

(a) (b)

Figure 6. Posterior distributions inferred from T1–T6 for (a) Case 1 (independent-parameter) and (b) Case 2 (correlated parameter).

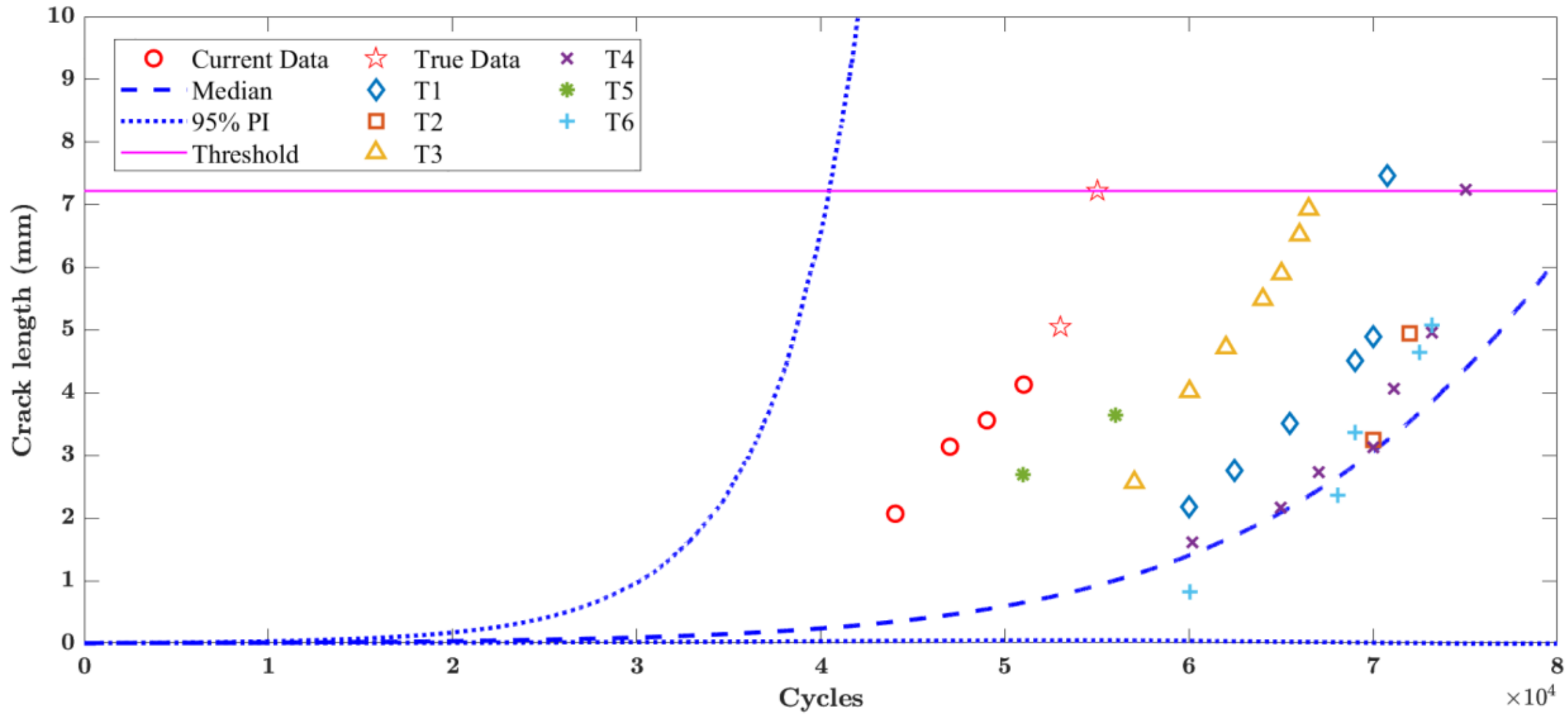

(a)

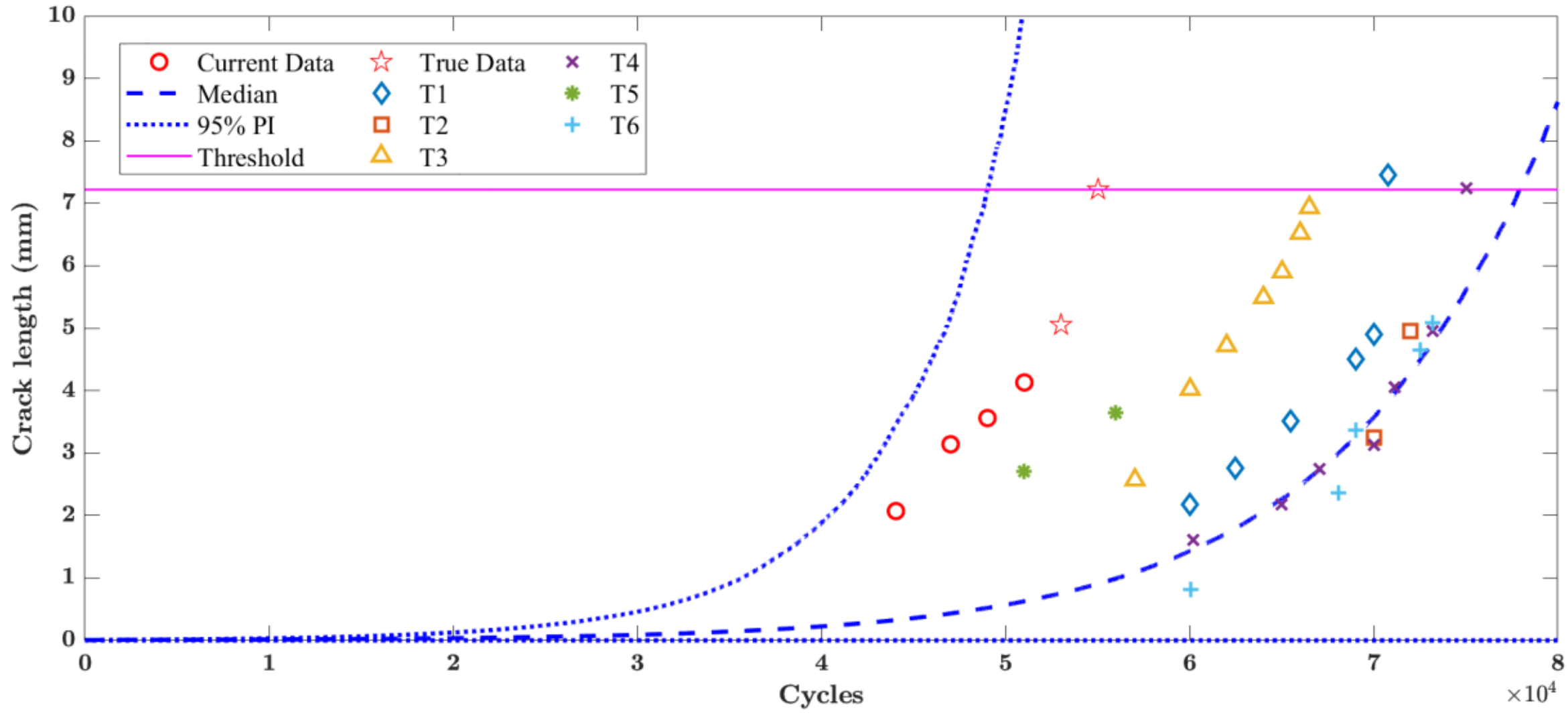

(b)

Figure 7. Predicted crack length using posterior samples of model parameters from Figure 6 for (a) Case 1: without correlation (b) Case 2: with correlation

### *3.3.3 Crack growth predictions for the current component (posterior updating with current data)*

To predict crack growth for the current component (T7), we update the historical posterior (obtained from T1–T6) using the available current measurements from T7 (red circles in Figure 7). This Bayesian updating step produces an updated posterior for the model parameters (Figure 8, summarized in Table 3), which is subsequently used to generate crack-growth predictions for the remainder of T7 (Figure 9). As expected, incorporating current data leads to narrower posterior distributions for the parameters and reduces predictive uncertainty in both cases. When comparing the two modeling assumptions, Case 2 (with correlation) consistently yields slightly smaller parameter uncertainty and narrower predictive credible intervals than Case 1.

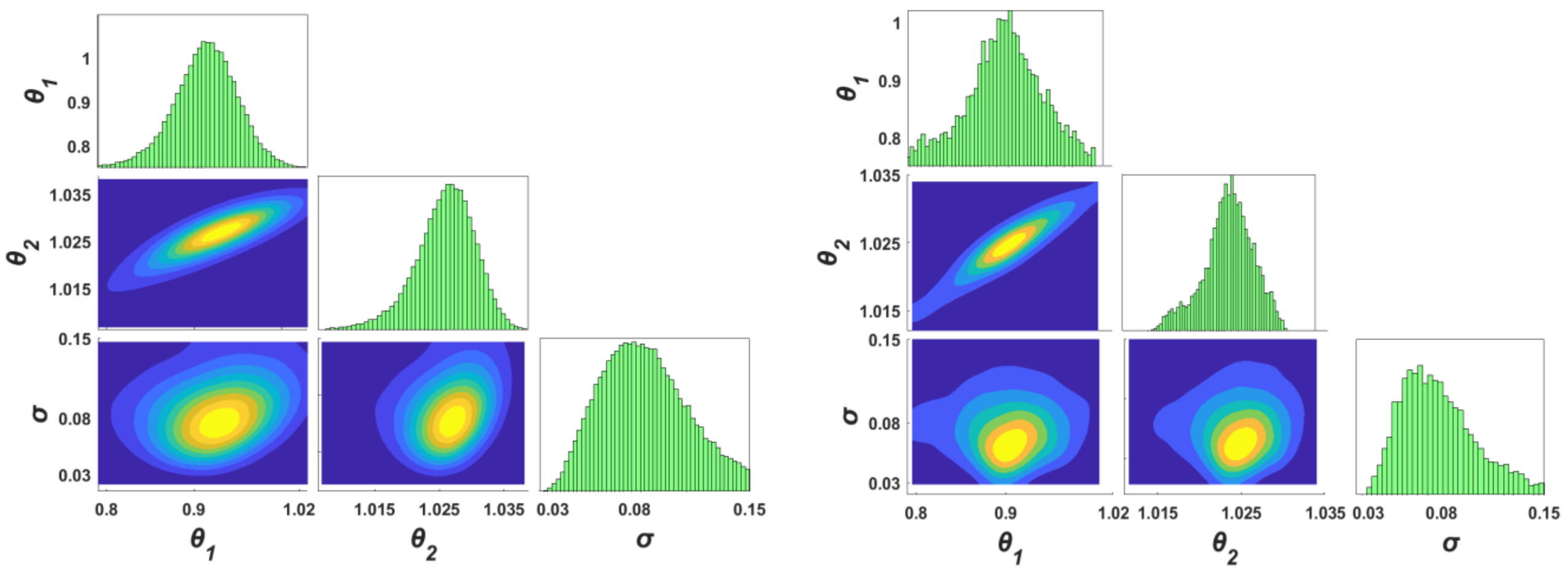

(a) (b)

Figure 8. Updated posterior distribution of model parameters considering current data from T7 for (a) Case 1: without correlation and (b) Case 2: with correlation

Table 3. Estimates of the statistics of the model parameters for the two cases (with current data from T7)

| | Statistics | $\theta_1$ | $\theta_2$ | $\sigma$ |
|---|---|---|---|---|
| Case 1 | Mean | 0.90 | 1.03 | 0.09 |
| | S.D. | 0.05 | 0.004 | 0.03 |
| Case 2 | Mean | 0.90 | 1.02 | 0.08 |
| | S.D. | 0.04 | 0.004 | 0.03 |

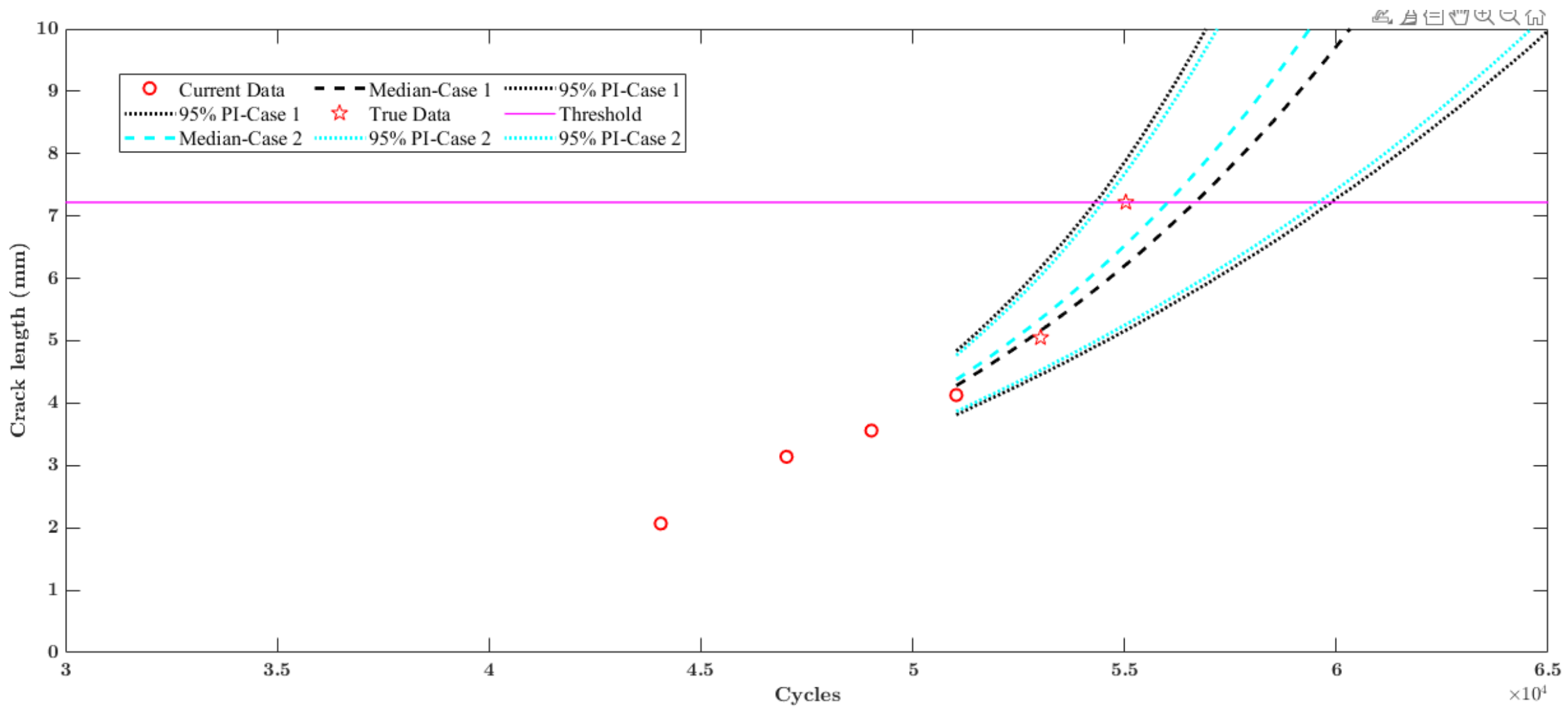

Figure 9. Predicted crack length using posterior samples of model parameters from Figure 8 using current data from T7 for Case 1 (without correlation) and Case 2 (with correlation).

### *3.3.4 Comparison with using a prior based on literature data*

The results presented above employ past experimental data to inform the prior distribution of model parameters. Alternatively, a prior could be selected based on information from the literature or design codes. To demonstrate this, we incorporate information from the Joint Committee on Structural Safety (JCSS) probabilistic model code to set up the prior for the model parameter log($C$) [53]. For the parameter $m$, we adopt a probabilistic model informed by the literature [54], which both refer to fatigue crack growth parameters for the same material as considered in this study. Based on this information, the prior distributions for the two parameters are defined as $m$~$N$(2.89, 0.29) and log($C$)~$N$(- 10.78,0.17), called literature prior in the following. The data of T7 is subsequently used to update prior information and to obtain the posterior distributions of the model parameters. The posterior results are illustrated in Figure 10(a), while the corresponding predictions of the crack length are presented in Figure 10(b). It is observed that when utilizing the literature prior, the mean values of the model parameters shift compared to those obtained using prior information from historical data. Although the predictions based on the literature prior produce larger uncertainty bounds for the crack length, these bounds fail to fully encompass the real measurements. This result confirms that using historical data within the HBM framework yields more accurate predictions.

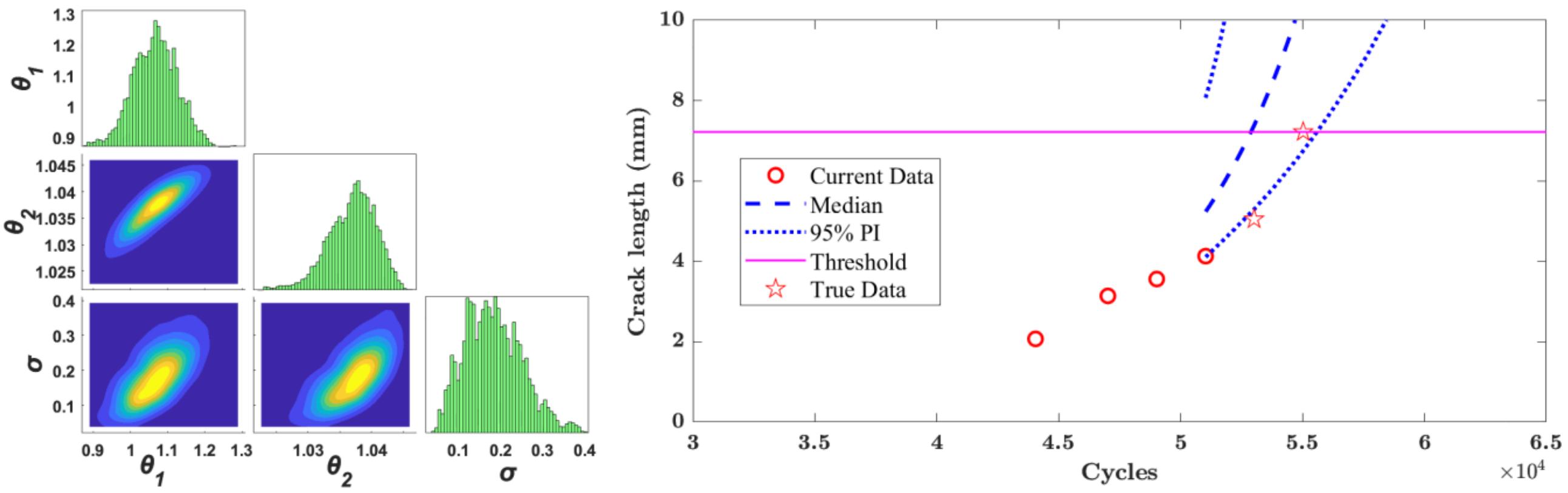

(a) (b)

Figure 10. (a) Posterior distribution of model parameters using data from T7, and (b) predicted crack length when using a prior from the literature instead of the historical data.

### *3.3.5 RUL prediction for T7 and T8*

The RUL can be readily computed from the crack length predictions. The RUL predictions for T7 using the historical data and the literature prior are first compared with the true RUL in Figure 11, where the true RUL is defined as the experimentally observed remaining life of T7 at the prediction time, which is 4001 cycles, and is indicated by the green marker in Figure 11. For this case, the RUL prediction obtained using the proposed framework aligns reasonably well with the true RUL, with the true value falling close to the mean of the predictive distribution. In contrast, the prediction obtained using the literature prior places the true RUL near the upper tail of the distribution, indicating a less accurate estimate for this experiment. In addition, the crack length prediction is also applied to T8, using the posterior samples of the hyperparameters shown in Figure 5. The corresponding RUL distributions are updated cycle by cycle as new crack-length measurements become available, and the results for T8 are shown in Figure 12. It is also seen that incorporating more current data progressively reduces the predictive uncertainty, as reflected by the narrowing RUL credible intervals. These results indicate that the proposed framework provides reasonable RUL predictions and effectively integrates new information as it becomes available.

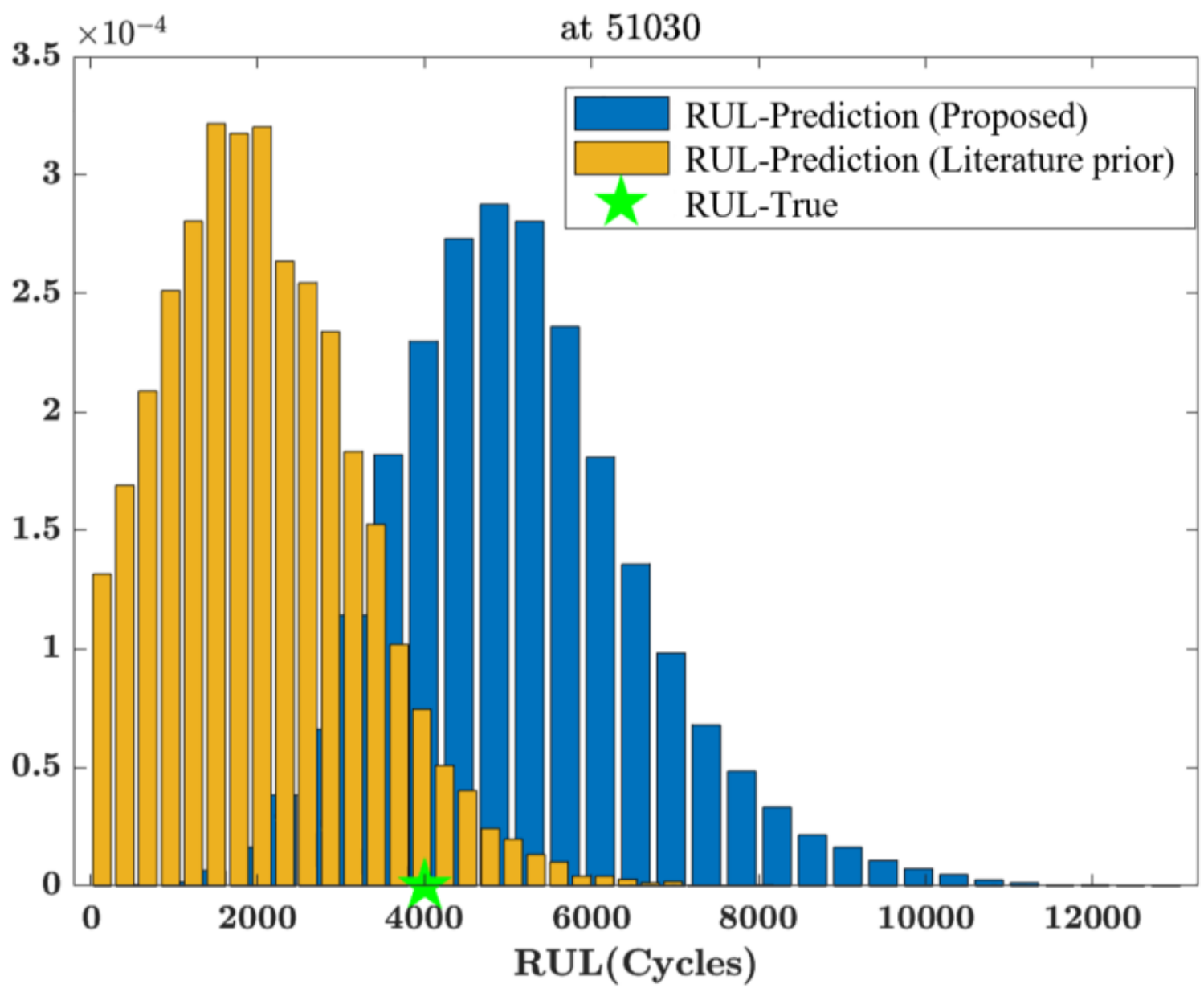

Figure 11. RUL prediction using both the proposed framework and the literature prior with current data from T7.

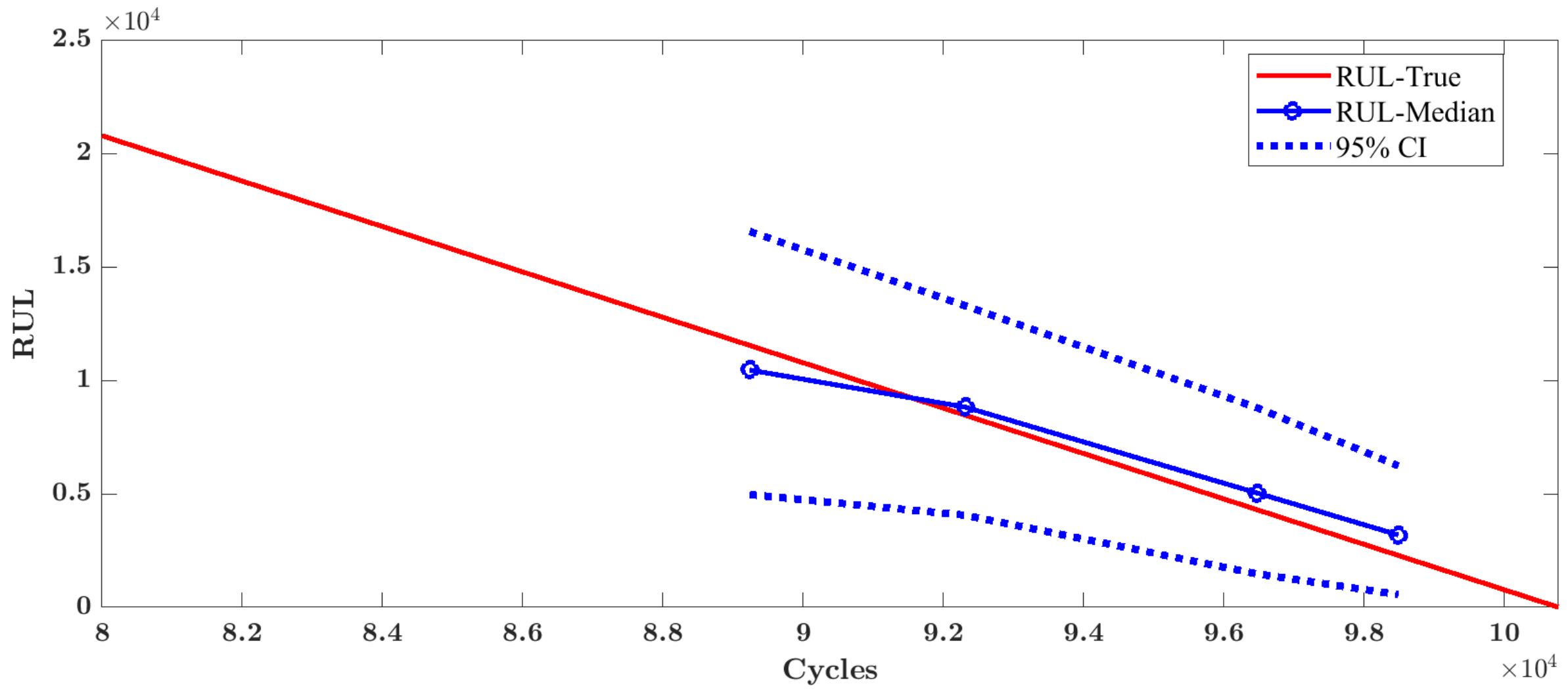

Figure 12. Predicted RUL distribution against the true RUL in function of cycles using current data from T8.

## 4. Application to Battery Degradation

### *4.1 Experimental datasets and battery degradation models*

The proposed framework is evaluated using battery data from the NASA Ames Research Center [55]. These data are widely used to investigate lithium-ion battery degradation prediction [3,35,56,57]. Specifically, the data used for analysis comes from four batteries called B0005, B0006, B0007 and B0018, shown in Figure 13. These batteries were tested under the same conditions at 23 $^{\circ}C$ to measure the degradation of their capacity. Repeated charging and discharging cycles resulted in accelerated aging of these batteries. Charging was conducted in a constant current mode at 1.5A until the battery voltage reached 4.2 V, and the discharging was carried out at a constant current level of 2A until the battery voltage fell to 2.7V for battery B0005, 2.5V for B0006, 2.2V for B0007 and 2.5V for B0018. The rated capacity of the tested batteries is 2Ahr, and the failure threshold is set to 1.4 Ahr. Datasets B0005–B0007 are treated as the historical datasets used to infer the historical posterior distribution of the hyperparameters, while dataset B0018 serves as the current component dataset on which the model is updated and predictions are performed.

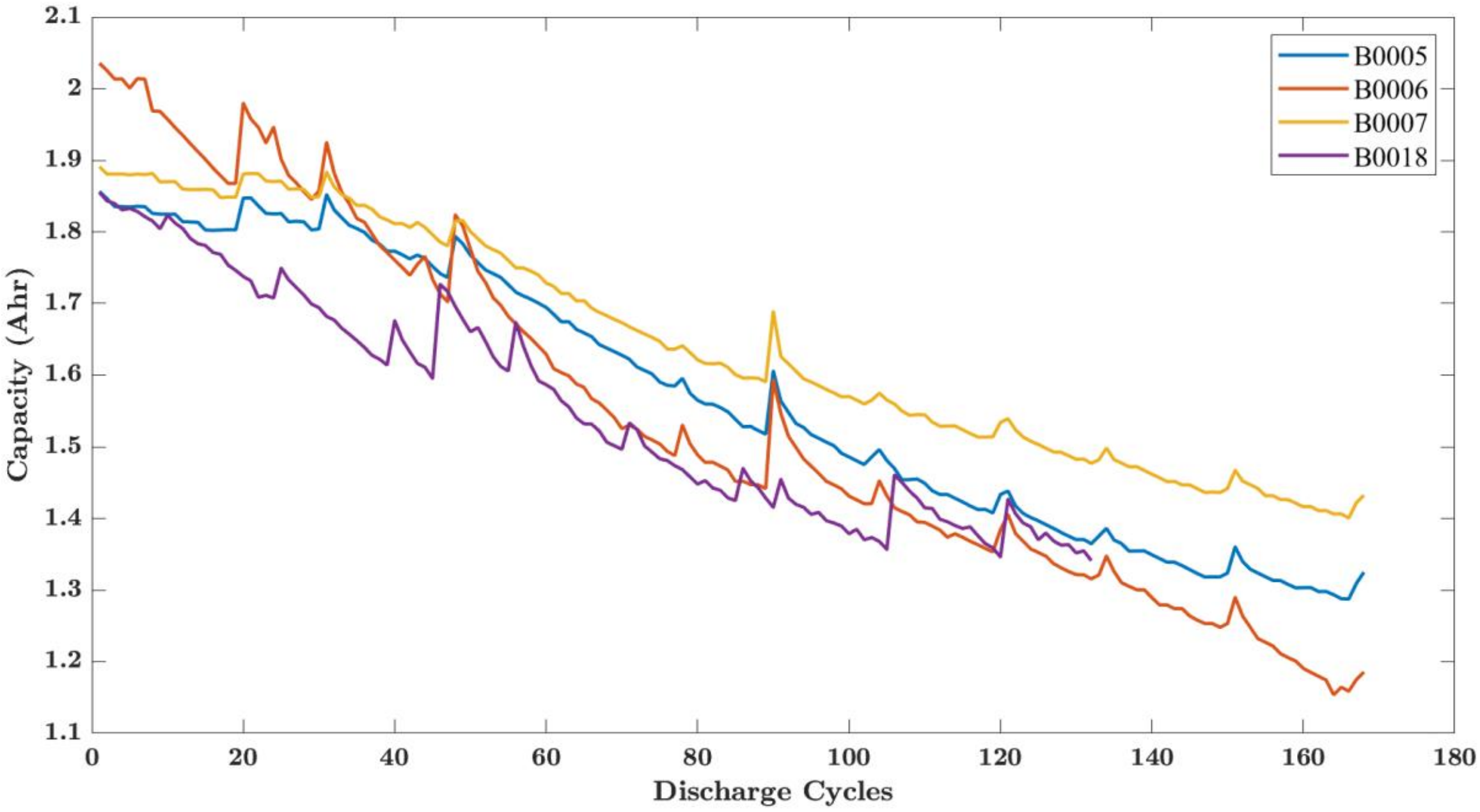

Figure 13. Battery degradation datasets provided by NASA Ames Research Center

Battery degradation models for this dataset have been explored in [3]. Therein, two heuristic degradation models are compared to describe the degradation behavior of the four lithium-ion batteries: the single exponential degradation model and the double exponential model. We investigate both models to determine which is more suitable for demonstrating of the HBM framework in the context of battery degradation. The single exponential degradation model is given by [3]:

$$Q = C_0 + a \bullet \exp\left(b / k\right), \tag{18}$$

where $Q$ is the capacity of the battery and $k$ denotes the discharge cycle, $C_0$ defines the initial capacity of the battery, $a$ and $b$ are model parameters that govern the capacity of the battery. Similar to the study in [3], the parameters $C_0$, $a$ and $b$ are estimated using the data. The double exponential degradation model is given by:

$$Q = \tilde{a} \bullet \exp\left(\tilde{b} \bullet k\right) + \tilde{c} \bullet \exp\left(\tilde{d} \bullet k\right), \tag{19}$$

where $\tilde{a}$, $\tilde{b}$, $\tilde{c}$, and $\tilde{d}$ are the model parameters that need to be estimated; parameters $\tilde{a}$ and $\tilde{c}$ characterize the initial capacity and are related to the internal impedance, while $\tilde{b}$ and $\tilde{d}$ represent the aging rate.

### *4.2 Setup of the numerical investigation*

The nominal values of the three parameters $C_{0,0}$, $a_0$ and $b_0$ for the single exponential model are chosen as 2, -1 and -100, while the nominal values of the parameters $\tilde{a}$, $\tilde{b}$, $\tilde{c}$, $\tilde{d}$ in the double exponential model are selected as 1.92, -0.02, -0.003 and -0.05. These values are selected based on the updated results of [3]. As in the fatigue crack example, the model parameters are expressed in normalized form to reduce scale differences and to improve numerical stability during inference. Therefore, the parameters in the single exponential model are assigned as $\theta_1 = C_0 / C_{0,0}$, $\theta_2 = a / a_0$ and $\theta_3 = b / b_0$. The parameters for the double exponential model $\boldsymbol{\theta} = [\theta_1, \theta_2, \theta_3, \theta_4]$ are defined in the same manner. The standard deviation $\sigma$ of the prediction error term is also estimated using Bayesian inference. The likelihood function for the $i$-th dataset at $k$-th cycle $y_{i,k}$ is given by a Gaussian PDF:

$$p\left(y_{i,k}\middle|\boldsymbol{\theta}_i\right)=\frac{1}{\sqrt{2\pi}\sigma}\exp[-\frac{(y_{i,k}-Q(\boldsymbol{\theta}_i,k))^2}{2\sigma^2}]. \tag{20}$$

### *4.3 Results and discussions*

### *4.3.1 Model comparisons and posterior distribution of hyperparameters*

We compare the performances of the single exponential and double exponential models, using a model selection procedure. To this end, we assess the log-evidence computed together with the posterior distribution of the hyperparameters through the computational procedure in Appendix A. The log-evidence values for single and double exponential models, using the HBM framework, are computed as 9.68 and 10.72 with the Transitional Markov Chain Monte Carlo (TMCMC) algorithm [58]. The double exponential model thus yields higher log-evidence values, indicating a better fit to the battery degradation data. Comparisons were also conducted between models with and without correlation parameters, suggesting that they do not significantly affect model predictions in the battery degradation case. Consequently, we employ the double exponential model with diagonal covariance matrix, with only the hyper standard deviations to be estimated. The hyperparameters $\boldsymbol{\mu}_{\boldsymbol{\theta}}$ and $\boldsymbol{\Sigma}_{\boldsymbol{\theta}}$ have uniform priors ranging from 0 to 1.8 for the hyper means and 0 to 0.4 for the hyper standard deviations. The historical posterior distributions of these hyperparameters, learned with the data from batteries B0005-B0007 are presented in Figure 14. These historical posterior samples are used for predicting both the model parameters and the battery capacity in the next section.

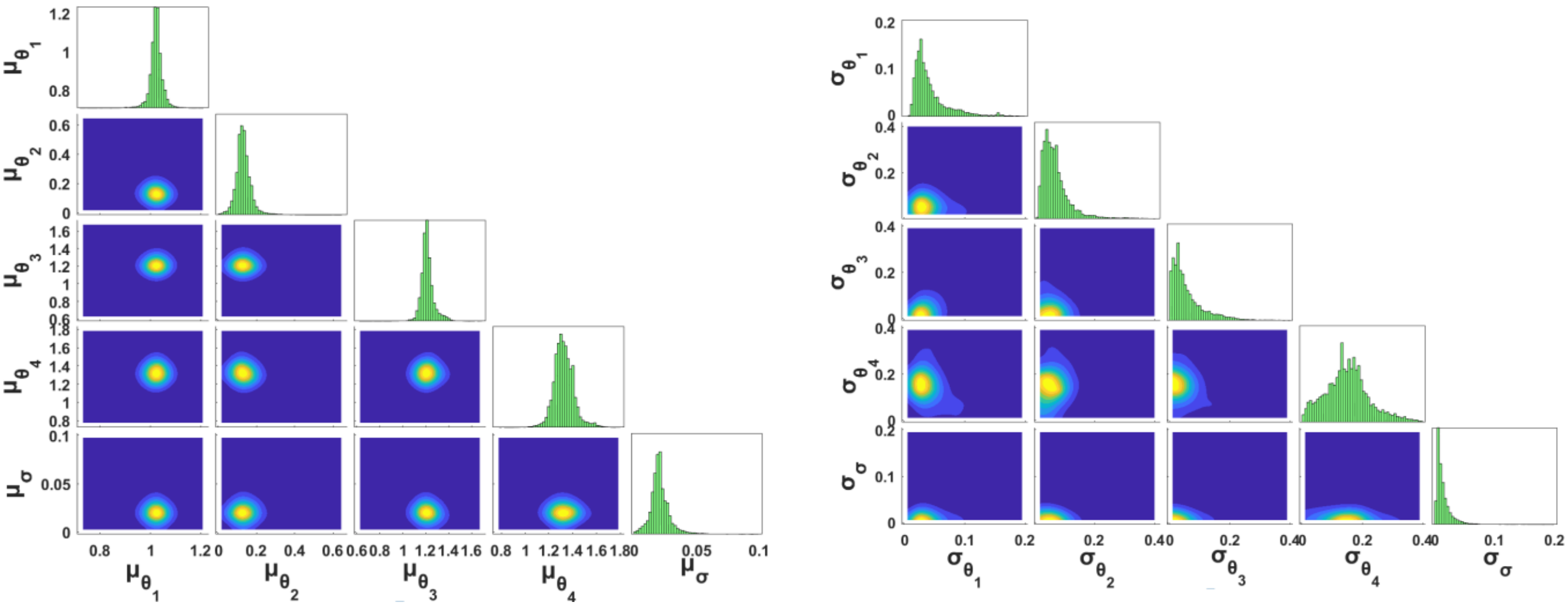

(a) Hyper mean (b) Hyper standard deviation

Figure 14. Historical posterior distribution of hyperparameters

### *4.2.2 Posterior distribution of the model parameters*

The posterior distributions of the model parameters are computed using either only historical data (from batteries B0005-B0007) or both the historical data to inform the parameter prior and the current data (from B0008) up to 70-th cycle for updating. The results of the posterior distributions are shown in Figure 15. With the incorporation of the current data, the updated posterior samples of the model parameters become more concentrated, indicating a reduction in uncertainty as the model parameters are refined using the updated information. Despite this improvement, a 2% model error persists, which can be attributed to measurement noise and the inherent limitations of the model. By incorporating the current data, narrower uncertainty bounds for the predictions are anticipated.

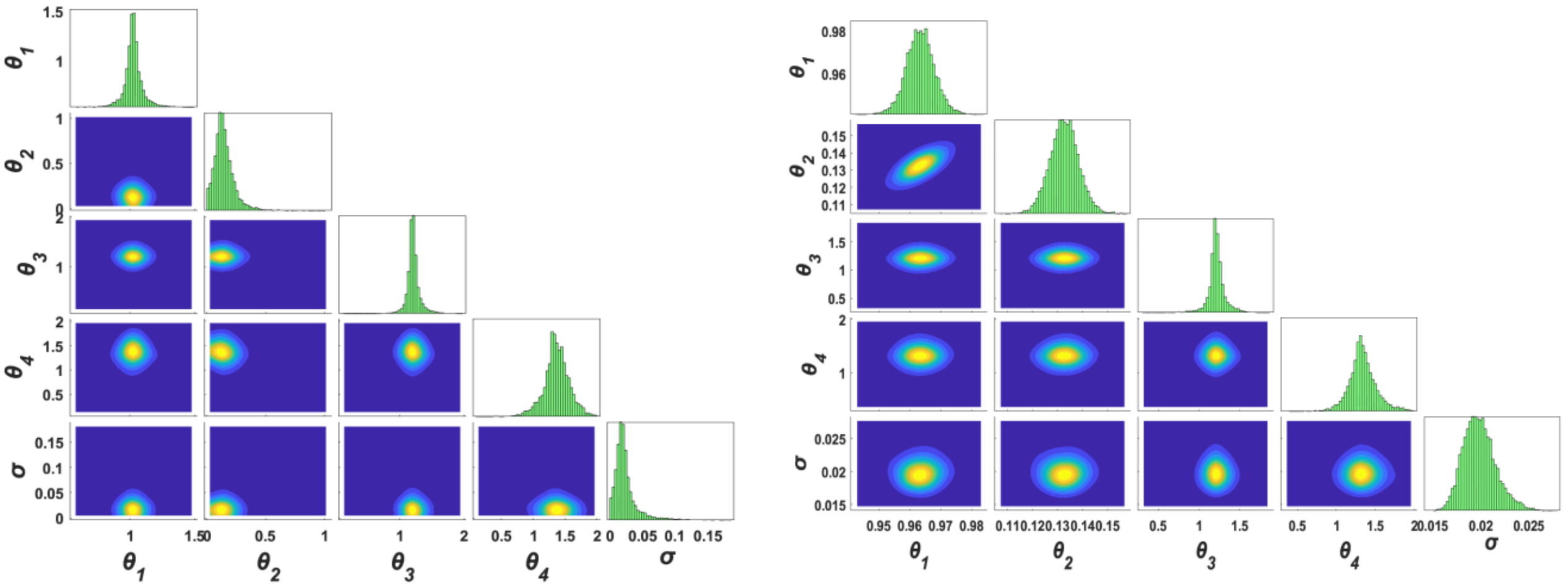

(a) (b)

Figure 15. Posterior distribution of the model parameters using (a) historical information only and (b) using also current information

#### *4.2.3 The capacity and RUL predictions*

The posterior samples of the model parameters shown in Figure 15 are used to predict the battery capacity. These predictions, along with their corresponding uncertainty bounds, are shown in Figure 16 for historical information only and in Figure 17 for historical information plus current data. Large uncertainty bounds are obtained when using only the historical datasets. As expected, including current data leads to significantly narrower uncertainty bounds. Once the predicted battery capacity is obtained, the RUL prediction is obtained through Eq. (12). The RUL predictions are plotted as a function of cycles in Figure 18, with data points shown every 2 cyclesFigure 18. To generate these predictions, the model's posterior distributions are computed using the current data for each two-cycle interval. It is observed that when using limited data, particularly from the 20$^{th}$ to the 55$^{th}$ cycle, the uncertainty bounds of the RUL prediction fluctuate significantly, and the true RUL lies outside the 95% credible interval of the prediction for a significant period. However, as more data is incorporated, the uncertainty bounds become narrower and the mean RUL prediction increasingly aligns with the true RUL. It is also notable that the measured capacity data exhibit irregular fluctuations. Such fluctuations likely arise from measurement noise, operational variability, or short-term electrochemical effects that are not captured by the parametric model, resulting in a model–data mismatch. This mismatch is absorbed by the additive model error term, which is approximately 2% in this case (Figure 15(b)) and accounts for the portion of the variability that the parametric model is not intended to capture.

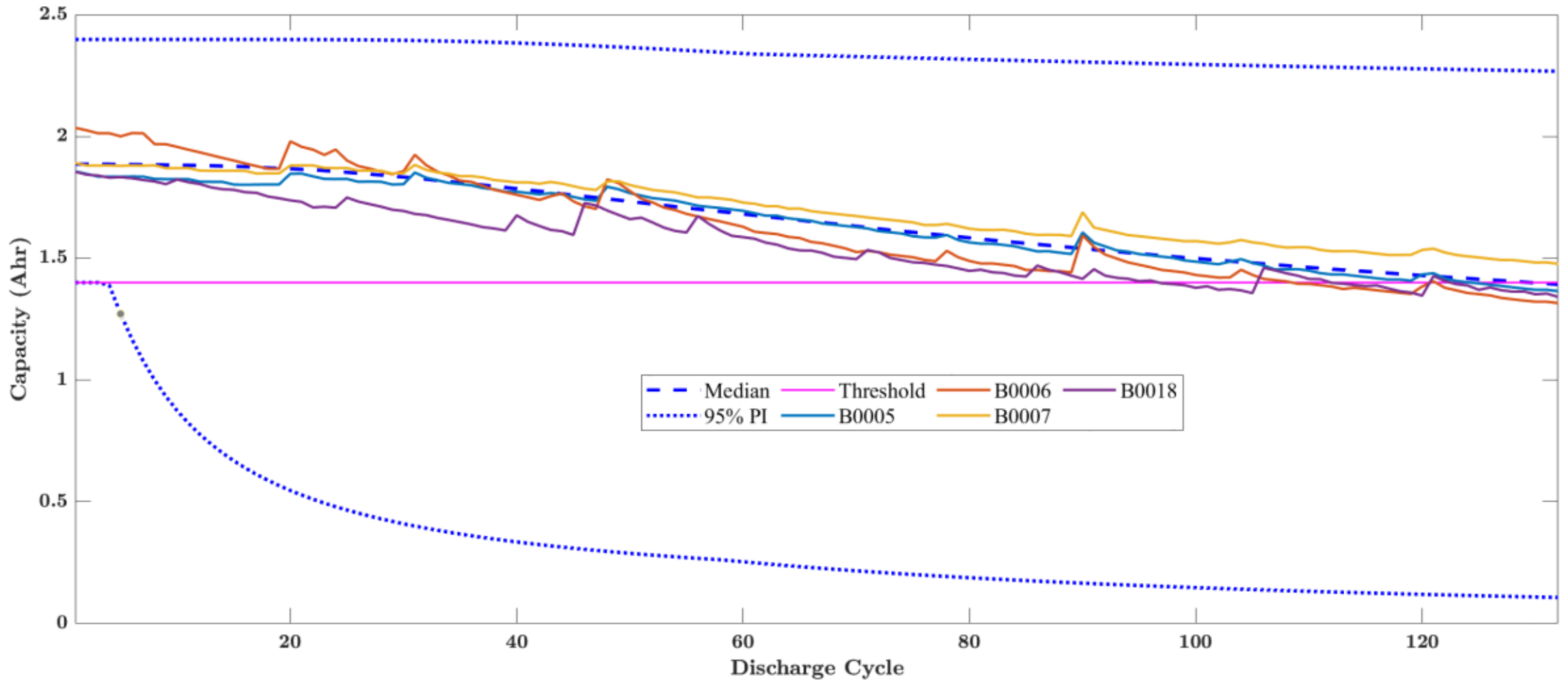

Figure 16. The capacity predictions using historical data only

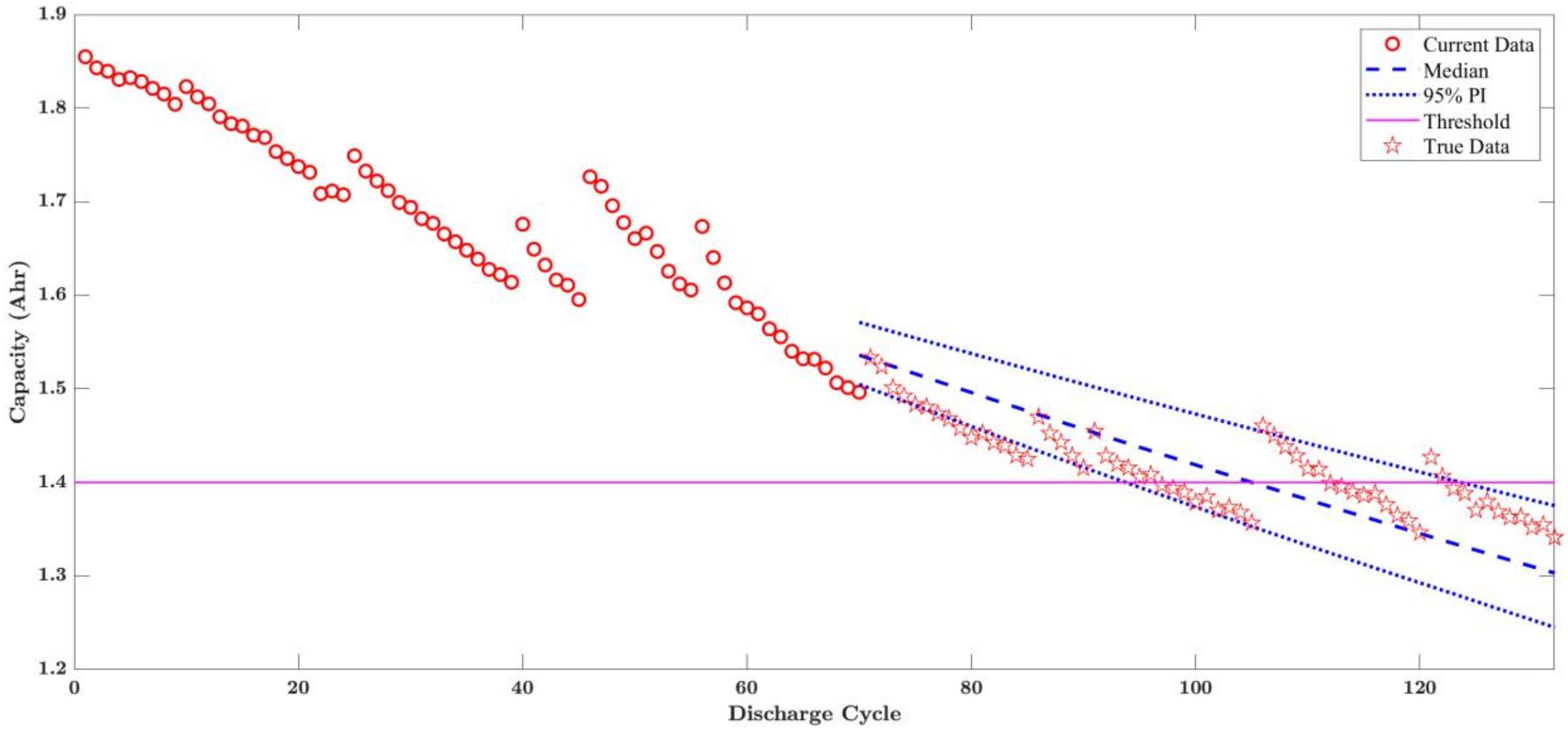

Figure 17. The capacity predictions with current information

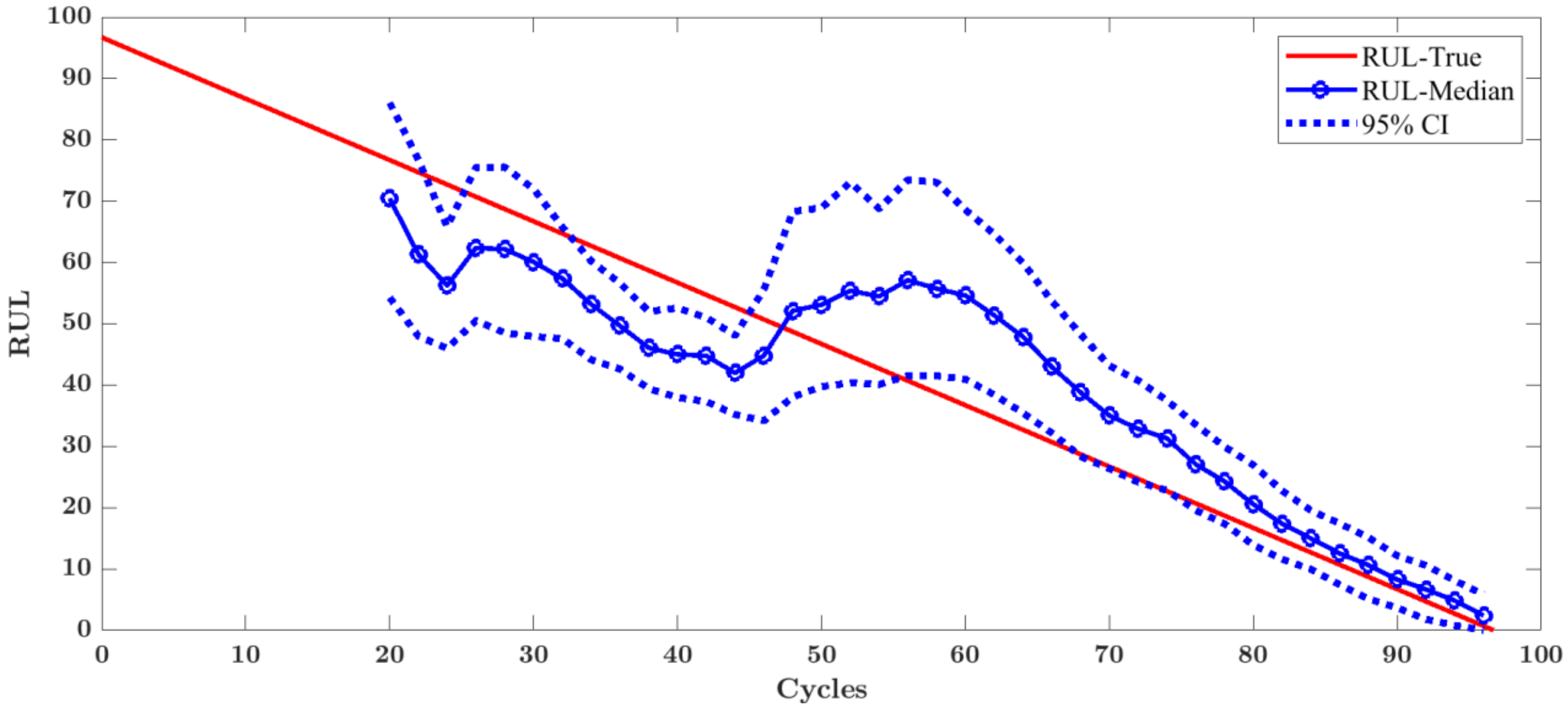

Figure 18. The RUL predictions in function of cycles

## 5. Conclusion

We propose a Hierarchical Bayesian Modelling (HBM) framework for Prognostics and Health Management (PHM). The framework addresses limitations in current approaches by integrating historical data and current operational data to provide more reliable and accurate Remaining Useful Life (RUL) predictions. The framework was investigated with two experimental cases with crack growth and lithium battery degradation. The results demonstrate that incorporating historical data through hyperparameter-informed priors enhances the predictive capabilities of the model, providing more accurate RUL predictions and managing uncertainty effectively. By leveraging an informed prior based on historical data from similar components for Bayesian updating with current component-specific data, the proposed HBM framework offers a robust and adaptive method for capturing the degradation process and its variability. It was also demonstrated that integrating correlation between model parameters as additional hyperparameters has the potential to refine model predictions, but not always.

Future work can extend this framework to multi-level systems, such as an aircraft engine, which comprises subsystems (e.g., turbofan, combustion chamber, lubrication system) and individual components (e.g., fan blades, fuel nozzles, bearings). This will enable the framework to make more comprehensive and reasonable predictions of system-level degradation and RUL, ensuring its applicability to complex, real-world scenarios.

**Acknowledgment:** Support by the Alexander von Humboldt Foundation is gratefully acknowledged.

**Appendix A.**

Algorithm: Compute the posterior distribution of the hyperparameters

1) **Step 1:** Model inference for each data set

**For** $i=1$ **to** $N$

Obtain samples $\boldsymbol{\theta}_i^{(k)}, k=1,2,\cdots,N_k$, from the posterior $p\left(\boldsymbol{\theta}_i \mid \mathbf{D}_i\right)$

[Using any Bayesian sampling algorithm]

**End**

**Store all samples of** $\boldsymbol{\theta}_i^{(k)}$

2) **Step 2:** Obtain samples of hyperparameters from the distribution

$$p\left(\boldsymbol{\psi} \middle| \mathbf{D}_h\right) \propto p\left(\boldsymbol{\mu}_{\boldsymbol{\theta}}, \boldsymbol{\Sigma}_{\boldsymbol{\theta}}\right) \prod_{i=1}^{N} \sum_{k=1}^{N_k} N\left(\boldsymbol{\theta}_i^{(k)} \mid \boldsymbol{\mu}_{\boldsymbol{\theta}}, \boldsymbol{\Sigma}_{\boldsymbol{\theta}}\right)$$

**For** $i=1$ **to** $N$

**For** $k=1$ **to** $N_s$

2.1) Take samples $\boldsymbol{\theta}_i^{(k)}$ from the first step

2.2) Construct function $p\left(\boldsymbol{\theta}_i^{(k)} \mid \boldsymbol{\psi}\right) = N\left(\boldsymbol{\theta}_i^{(k)} \mid \boldsymbol{\mu}_{\boldsymbol{\theta}}, \boldsymbol{\Sigma}_{\boldsymbol{\theta}}\right)$

**End**

2.3) **Sum:** $\sum_{k=1}^{N_k} N\left(\boldsymbol{\theta}_i^{(k)} \mid \boldsymbol{\mu}_{\boldsymbol{\theta}}, \boldsymbol{\Sigma}_{\boldsymbol{\theta}}\right)$

**End**

2.4) **Prod:** $\prod_{i=1}^{N} \sum_{k=1}^{N_k} N\left(\boldsymbol{\theta}_i^{(k)} \mid \boldsymbol{\mu}_{\boldsymbol{\theta}}, \boldsymbol{\Sigma}_{\boldsymbol{\theta}}\right)$

## REFERENCES

[1] Y. Hu, X. Miao, Y. Si, E. Pan, E. Zio, Prognostics and health management : A review from the perspectives of design , development and decision, Reliab. Eng. Syst. Saf. 217 (2022) 108063. https://doi.org/10.1016/j.ress.2021.108063.

[2] E. Zio, Prognostics and Health Management (PHM): Where are we and where do we (need to) go in theory and practice, Reliab. Eng. Syst. Saf. 218 (2022) 108119. https://doi.org/10.1016/j.ress.2021.108119.

[3] M. Catelani, L. Ciani, S. Member, R. Fantacci, G. Patrizi, S. Member, B. Picano, Remaining Useful Life Estimation for Prognostics of Lithium-Ion Batteries Based on Recurrent Neural Network, IEEE Trans. Instrum. Meas. 70 (2021) 1–11. https://doi.org/10.1109/TIM.2021.3111009.

[4] J. Chen, S. Yuan, L. Qiu, H. Wang, W. Yang, On-line prognosis of fatigue crack propagation based on Gaussian weight-mixture proposal particle filter, Ultrasonics 82 (2018) 134–144. https://doi.org/10.1016/j.ultras.2017.07.016.

[5] H. Meng, M. Geng, T. Han, Long short-term memory network with Bayesian optimization for health prognostics of lithium-ion batteries based on partial incremental capacity analysis, Reliab.

Eng. Syst. Saf. 236 (2023) 109288. https://doi.org/10.1016/j.ress.2023.109288.
[6] B. Ospina Agudelo, W. Zamboni, F. Postiglione, E. Monmasson, Battery State-of-Health estimation based on multiple charge and discharge features, Energy 263 (2023) 125637. https://doi.org/10.1016/j.energy.2022.125637.
[7] M. Berecibar, F. Devriendt, M. Dubarry, I. Villarreal, N. Omar, W. Verbeke, J. Van Mierlo, Online state of health estimation on NMC cells based on predictive analytics, J. Power Sources 320 (2016) 239–250. https://doi.org/10.1016/j.jpowsour.2016.04.109.
[8] S. Pfingstl, M. Zimmermann, On integrating prior knowledge into Gaussian processes for prognostic health monitoring, Mech. Syst. Signal Process. 171 (2022) 108917. https://doi.org/10.1016/j.ymssp.2022.108917.
[9] S.M.M. Seyed, X. Jin, J. Ni, Physics-based Gaussian process for the health monitoring for a rolling bearing, Acta Astronaut. 154 (2019) 133–139. https://doi.org/10.1016/j.actaastro.2018.10.029.
[10] L. Jing, M. Zhao, P. Li, X. Xu, A convolutional neural network based feature learning and fault diagnosis method for the condition monitoring of gearbox, Meas. J. Int. Meas. Confed. 111 (2017) 1–10. https://doi.org/10.1016/j.measurement.2017.07.017.
[11] S. Shen, M. Sadoughi, X. Chen, M. Hong, C. Hu, A deep learning method for online capacity estimation of lithium-ion batteries, J. Energy Storage 25 (2019) 100817. https://doi.org/10.1016/j.est.2019.100817.
[12] S. Khan, T. Yairi, A review on the application of deep learning in system health management, Mech. Syst. Signal Process. 107 (2018) 241–265. https://doi.org/10.1016/j.ymssp.2017.11.024.
[13] M. Raissi, P. Perdikaris, G.E. Karniadakis, Physics-informed neural networks: A deep learning framework for solving forward and inverse problems involving nonlinear partial differential equations, J. Comput. Phys. 378 (2019) 686–707. https://doi.org/10.1016/j.jcp.2018.10.045.
[14] A. Thelen, Y. Hui, S. Shen, S. Laflamme, S. Hu, H. Ye, C. Hu, Integrating physics-based modeling and machine learning for degradation diagnostics of lithium-ion batteries, Energy Storage Mater. 50 (2022) 668–695. https://doi.org/10.1016/j.ensm.2022.05.047.
[15] A. Subramanian, S. Mahadevan, Probabilistic physics-informed machine learning for dynamic systems, Reliab. Eng. Syst. Saf. 230 (2023) 108899. https://doi.org/10.1016/j.ress.2022.108899.
[16] S. Kim, J.H. Choi, N.H. Kim, Data-driven prognostics with low-fidelity physical information for digital twin: physics-informed neural network, Struct. Multidiscip. Optim. 65 (2022) 1–16. https://doi.org/10.1007/s00158-022-03348-0.
[17] Y. Xu, S. Kohtz, J. Boakye, P. Gardoni, P. Wang, Physics-informed machine learning for reliability and systems safety applications: State of the art and challenges, Reliab. Eng. Syst. Saf. 230 (2023) 108900. https://doi.org/10.1016/j.ress.2022.108900.
[18] S. Navidi, A. Thelen, T. Li, C. Hu, Physics-informed machine learning for battery degradation diagnostics : A comparison of state-of-the-art methods, Energy Storage Mater. 68 (2024) 103343. https://doi.org/10.1016/j.ensm.2024.103343.
[19] V. Nemani, L. Biggio, X. Huan, Z. Hu, O. Fink, A. Tran, Uncertainty quantification in machine learning for engineering design and health prognostics : A tutorial, Mech. Syst. Signal Process. 205 (2023) 110796. https://doi.org/10.1016/j.ymssp.2023.110796.
[20] Y. Hui, M. Li, A. Downey, S. Shen, V. Pavan, H. Ye, C. Vanelzen, G. Jain, S. Hu, S. Laflamme, C. Hu, Physics-based prognostics of implantable-grade lithium-ion battery for remaining useful life prediction, J. Power Sources 485 (2021) 229327.

https://doi.org/10.1016/j.jpowsour.2020.229327.
[21] J. Liu, D. Wang, J.Z. Kong, N. Li, Z. Peng, K.L. Tsui, New Look at Bayesian Prognostic Methods, IEEE Trans. Autom. Sci. Eng. PP (2024) 1–16. https://doi.org/10.1109/TASE.2024.3474790.
[22] E. Zio, G. Peloni, Particle filtering prognostic estimation of the remaining useful life of nonlinear components, Reliab. Eng. Syst. Saf. 96 (2011) 403–409. https://doi.org/10.1016/j.ress.2010.08.009.
[23] D. An, J. Choi, T.L. Schmitz, N.H. Kim, In situ monitoring and prediction of progressive joint wear using Bayesian statistics, Wear 270 (2011) 828–838. https://doi.org/10.1016/j.wear.2011.02.010.
[24] H. Jun, N.H. Kim, J. Choi, A robust health prediction using Bayesian approach guided by physical constraints, Reliab. Eng. Syst. Saf. 244 (2024) 109954. https://doi.org/10.1016/j.ress.2024.109954.
[25] I. Behmanesh, B. Moaveni, G. Lombaert, C. Papadimitriou, Hierarchical Bayesian model updating for structural identification, Mech. Syst. Signal Process. 64 (2015) 360–376.
[26] M. Song, B. Moaveni, C. Papadimitriou, A. Stavridis, Accounting for amplitude of excitation in model updating through a hierarchical Bayesian approach: Application to a two-story reinforced concrete building, Mech. Syst. Signal Process. 123 (2019) 68–83. https://doi.org/10.1016/j.ymssp.2018.12.049.
[27] X. Jia, O. Sedehi, C. Papadimitriou, L.S. Katafygiotis, B. Moaveni, Hierarchical Bayesian modeling framework for model updating and robust predictions in structural dynamics using modal features, Mech. Syst. Signal Process. 170 (2022) 108784. https://doi.org/10.1016/j.ymssp.2021.108784.
[28] J. Ching, S. Wu, K.-K. Phoon, Constructing Quasi-Site-Specific Multivariate Probability Distribution Using Hierarchical Bayesian Model, J. Eng. Mech. 147 (2021) 1–18. https://doi.org/10.1061/(asce)em.1943-7889.0001964.
[29] S. Wu, J. Ching, K.K. Phoon, Quasi-site-specific soil property prediction using a cluster-based hierarchical Bayesian model, Struct. Saf. 99 (2022) 102253. https://doi.org/10.1016/j.strusafe.2022.102253.
[30] M. Tabarroki, J. Ching, S.H. Yuan, K.K. Phoon, F. Teng, Data-driven hierarchical Bayesian model for predicting wall deflections in deep excavations in clay, Comput. Geotech. 168 (2024) 106135. https://doi.org/10.1016/j.compgeo.2024.106135.
[31] S. Wu, P. Angelikopoulos, G. Tauriello, C. Papadimitriou, P. Koumoutsakos, Fusing heterogeneous data for the calibration of molecular dynamics force fields using hierarchical Bayesian models, J. Chem. Phys. 145 (2016). https://doi.org/10.1063/1.4967956.
[32] A. Economides, G. Arampatzis, D. Alexeev, S. Litvinov, L. Amoudruz, L. Kulakova, C. Papadimitriou, P. Koumoutsakos, Hierarchical Bayesian Uncertainty Quantification for a Model of the Red Blood Cell, Phys. Rev. Appl. 15 (2021). https://doi.org/10.1103/PhysRevApplied.15.034062.
[33] S. Wu, P. Angelikopoulos, C. Papadimitriou, R. Moser, P. Koumoutsakos, A hierarchical Bayesian framework for force field selection in molecular dynamics simulations, Philos. Trans. R. Soc. A Math. Phys. Eng. Sci. 374 (2016). https://doi.org/10.1098/rsta.2015.0032.
[34] X. Xu, Z. Li, N. Chen, A Hierarchical Model for Lithium-Ion Battery Degradation Prediction, IEEE Trans. Reliab. 65 (2016) 310–325. https://doi.org/10.1109/TR.2015.2451074.

[35] M. Mishra, J. Martinsson, M. Rantatalo, K. Goebel, Bayesian hierarchical model-based prognostics for lithium-ion batteries, Reliab. Eng. Syst. Saf. 172 (2018) 25–35. https://doi.org/10.1016/j.ress.2017.11.020.

[36] X. Jia, O. Sedehi, C. Papadimitriou, L.S. Katafygiotis, B. Moaveni, Nonlinear model updating through a hierarchical Bayesian modeling framework, Comput. Methods Appl. Mech. Eng. 392 (2022) 114646. https://doi.org/10.1016/j.cma.2022.114646.

[37] X. Jia, W.J. Yan, C. Papadimitriou, K.V. Yuen, An analytically tractable solution for hierarchical Bayesian model updating with variational inference scheme, Mech. Syst. Signal Process. 189 (2023) 110060. https://doi.org/10.1016/j.ymssp.2022.110060.

[38] C.E. Shannon, A mathematical theory of communication, Bell Syst. Tech. J. 27 (1948) 379–423. 10.1002/j.1538-7305.1948.tb01338.x.

[39] S. Wu, P. Angelikopoulos, J.L. Beck, P. Koumoutsakos, Hierarchical Stochastic Model in Bayesian Inference for Engineering Applications: Theoretical Implications and Efficient Approximation, ASCE-ASME J. Risk Uncertain. Eng. Syst. Part B Mech. Eng. 5 (2019). https://doi.org/10.1115/1.4040571.

[40] C. Papadimitriou, J.L. Beck, L.S. Katafygiotis, Asymptotic expansions for reliability and moments of uncertain systems, J. Eng. Mech. 123 (1997) 1219–1229.

[41] D. Patsialis, A.P. Kyprioti, A.A. Taflanidis, Bayesian calibration of hysteretic reduced order structural models for earthquake engineering applications, Eng. Struct. 224 (2020) 111204. https://doi.org/10.1016/j.engstruct.2020.111204.

[42] S. Kim, J.H. Choi, N.H. Kim, Inspection schedule for prognostics with uncertainty management, Reliab. Eng. Syst. Saf. 222 (2022) 108391. https://doi.org/10.1016/j.ress.2022.108391.

[43] I. De Pater, A. Reijns, M. Mitici, Alarm-based predictive maintenance scheduling for aircraft engines with imperfect Remaining Useful Life prognostics, Reliab. Eng. Syst. Saf. 221 (2022) 108341. https://doi.org/10.1016/j.ress.2022.108341.

[44] D. Koutas, D. Straub, Leaf it to renewal: Improved predictive maintenance policies via renewal theory and decision trees, (2025). http://arxiv.org/abs/2509.20145.

[45] J. He, X. Guan, T. Peng, Y. Liu, A multi-feature integration method for fatigue crack detection and crack length estimation in riveted lap joints using Lamb waves, Smart Mater. Struct. (2019). https://doi.org/10.1088/0964-1726/22/10/105007.

[46] T. Peng, J. He, Y. Xiang, Y. Liu, A. Saxena, J. Celaya, K. Goebel, Probabilistic fatigue damage prognosis of lap joint using Bayesian updating, J. Intell. Mater. Syst. Struct. 26 (2015) 965–979. https://doi.org/10.1177/1045389X14538328.

[47] J. Yang, J. He, X. Guan, D. Wang, H. Chen, A probabilistic crack size quantification method using in-situ Lamb wave test and Bayesian updating, Mech. Syst. Signal Process. 78 (2016) 118–133. https://doi.org/10.1016/j.ymssp.2015.06.017.

[48] M. Rao, X. Yang, D. Wei, Y. Chen, L. Meng, M.J. Zuo, Structure Fatigue Crack Length Estimation and Prediction Using Ultrasonic Wave Data Based on Ensemble Linear Regression and Paris ' s Law, Int. J. Progn. Heal. Manag. (2020) 1–14.

[49] S. Ojha, A. Shelke, Probabilistic Deep Learning Approach for Fatigue Crack Width Estimation and Prognosis in Lap Joint Using Acoustic Waves, J. Nondestruct. Eval. Diagnostics Progn. Eng. Syst. 8 (2025) 1–16. https://doi.org/10.1115/1.4065550.

[50] Y. Xiang, Y. Liu, An equivalent stress level model for efficient fatigue crack growth prediction, Collect. Tech. Pap. - AIAA/ASME/ASCE/AHS/ASC Struct. Struct. Dyn. Mater. Conf. (2011).

https://doi.org/10.2514/6.2011-2033.
[51] N.-H. Kim, D. An, J.-H. Choi, Prognostics and Health Management of Systems, 2017. https://doi.org/10.1007/978-3-319-44742-1.
[52] O.B. Downs, Slice sampling: Discussion, Ann. Stat. 31 (2003) 743–748.
[53] Joint Committee Structural Safety, JCSS Probabilistic Model Code: Resistance Models, (2013) 1–19.
[54] X.Y. Long, K. Liu, C. Jiang, Y. Xiao, S.C. Wu, Uncertainty propagation method for probabilistic fatigue crack growth life prediction, Theor. Appl. Fract. Mech. 103 (2019). https://doi.org/10.1016/j.tafmec.2019.102268.
[55] B. Saha, K. Goebel, Uncertainty management for diagnostics and prognostics of batteries using Bayesian techniques, IEEE Aerosp. Conf. Proc. (2008) 1–8. https://doi.org/10.1109/AERO.2008.4526631.
[56] X. Li, M. Lyu, K. Li, X. Gao, C. Liu, Z. Zhang, Lithium-ion battery state of health estimation based on multi-source health indicators extraction and sparse Bayesian learning, Energy 282 (2023) 128445. https://doi.org/10.1016/j.energy.2023.128445.
[57] D. Liu, Y. Luo, L. Guo, Y. Peng, Uncertainty Quantification of Fusion Prognostics for Lithium-ion Battery Remaining Useful Life Estimation, 2013 IEEE Conf. Progn. Heal. Manag. (2013) 1–8. https://doi.org/10.1109/ICPHM.2013.6621441.
[58] J. Ching, Y.-C. Chen, Transitional Markov chain Monte Carlo method for Bayesian model updating, model class selection, and model averaging, J. Eng. Mech. 133 (2007) 816–832. https://doi.org/10.1061/(ASCE)0733-9399(2007)133:7(816).